\documentclass[12pt]{iopart}
\pdfoutput=1 
\usepackage[utf8x]{inputenc}
\usepackage[UKenglish]{babel}
\expandafter\let\csname equation*\endcsname\relax
\expandafter\let\csname endequation*\endcsname\relax
\usepackage{amsmath}
\usepackage{iopams}
\usepackage{amssymb}
\usepackage{fullpage}
\usepackage{bm}
\usepackage{bbold}
\usepackage{appendix}
\usepackage{graphicx}
\usepackage{color}
\usepackage{tikz}
\usetikzlibrary{calc,decorations.markings}
\newcommand{\rhe}{\hat{R}_i^{\rm eff}(t,t')}
\newcommand{\hc}{B}
\newcommand{\be}{\begin{equation}}
\newcommand{\ee}{\end{equation}}

\newcommand{\ap}{\alpha}

\newcommand{\lhe}{l_i^{\rm eff}(t)}

\usepackage{cite}
\numberwithin{equation}{section}
\allowdisplaybreaks[1]
\begin{document}
\title[Extended Plefka Expansion with hidden nodes]{Inference for dynamics of continuous variables: the Extended Plefka Expansion with hidden nodes}
\author{B Bravi$^1$\footnote{Current affiliation: Institute of Theoretical Physics, Ecole Polytechnique F\'ed\'erale de Lausanne (EPFL), CH-1015 Lausanne, Switzerland}, 
 and P Sollich$^1$}
\address{$^1$ Department of Mathematics, King's College London,Strand, London, WC2R 2LS UK}
\ead{barbara.bravi@epfl.ch and peter.sollich@kcl.ac.uk}
\date{}

\begin{abstract}
We consider the problem of a subnetwork of observed nodes embedded into a larger bulk of unknown (i.e.\ hidden) nodes, where the aim is to infer these hidden states given information about the 
 subnetwork dynamics. The biochemical networks underlying many cellular and metabolic processes are important realizations of such a scenario as typically one is interested in reconstructing
 the time evolution of 
 unobserved chemical concentrations starting from the experimentally more accessible ones. 
We present an application to this problem of a novel dynamical mean field approximation, the Extended Plefka Expansion, 
 which is based on a path integral description of the stochastic dynamics. 
 As a paradigmatic model we study the stochastic linear dynamics of 
 continuous degrees of freedom interacting via random Gaussian couplings. 
 The resulting joint distribution is known to be Gaussian and this allows us to fully characterize
 the posterior statistics of the hidden nodes. In particular the equal-time hidden-to-hidden variance -- conditioned
 on observations -- gives the expected error at each node when the hidden time courses are predicted based on the observations. 
 We assess the accuracy of the Extended Plefka Expansion in predicting these single node variances as well as error correlations over time, 
 focussing on the role of the system size and the number of observed nodes.
\end{abstract}
\maketitle

\section{Introduction}

The problem of reconstructing the time evolution of a system given some measurements of its dynamics has seen much recent interest in the statistical physics 
community \cite{inference1,inference2,roudi1}. Given a temporal sequence of observed variables, the task is to infer the states of 
other variables that are not observed, based on knowledge of the interaction parameters. Techniques for tackling this problem could have a significant impact in 
systems biology, where dealing with missing variables and experimental limitations requires the development of novel inference frameworks, 
to enable quantitative modelling on the basis of experimental data \cite{sanguinetti,stumpf,confounding}.

In machine learning and pattern recognition, the problem of inference from data has been addressed using e.g.\ state space models \cite{bishop} that 
introduce ``hidden'' variables playing the role of unobserved 
states. The best-known example of this type is probably the Kalman filter \cite{kalmanor}, for the case of stochastic linear dynamics 
of continuous degrees of freedom (d.o.f.). Here we concentrate precisely on such a dynamics.
A wide range of microscopic biological processes can be captured qualitatively even by this simplified setting, such as 
inter- and intra-cellular biochemical networks where interactions parameters are given by reaction rates and one may be 
interested in analyzing fluctuations around a steady state \cite{LNA}.
The joint distribution over observed and hidden states is Gaussian in our model and this allows us to analyze the posterior statistics of the hidden dynamics. 
The second order statistics in particular tells us how the observed dynamics constrains the unknown hidden dynamics, with the posterior variances quantifying the 
degree of uncertainty in hidden state prediction.  

We tackle the inference problem for our setting by means of the Extended Plefka Expansion that we have recently developed \cite{bravi},
a dynamical mean field theory for continuous degrees of freedom. This is
general enough to allow us to treat a wide range of dynamical interactions, i.e.\ couplings of any symmetry, so that we can probe both equilibrium and non-equilibrium networks with only partially observed dynamics. This flexibility makes our approach in principle widely applicable to real data from systems driven out of equilibrium by fluxes, with biological networks being a case
in point as illustrated e.g.\ by studies on non-equilibrium steady states for reaction fluxes \cite{marinari} and 
``near" symmetry features in metabolic networks \cite{holme}.

Inference problems for non-equilibrium systems have already been investigated for neural data, using either two-state units (Ising spins) \cite{roudi1} or deterministic 
continuous-valued hidden units \cite{tyrcha}. These studies were motivated by modelling populations of neurons, and concentrated on finding learning rules, 
while here we study continuous d.o.f.\ with random linear interactions and our focus is on
a computationally efficient and accurate estimate of the state inference error, including
quantitative tests of its performance.
Note that for the overall prediction of dynamics in a system that is only partially observed, 
(interaction) parameter estimation is required in addition to state inference. We focus on the latter problem in this paper, which is in fact always part of parameter learning. 
For example, existing algorithmic tools such as Expectation Propagation \cite{minka,opperwinther}
iterate between estimating the hidden states (given the interactions) and estimating the interactions  
(given the updated states).

The paper is organized as follows. In section \ref{sec:pwoEPEHN} we set out our model in terms of a set of dynamical equations for observed and hidden continuous 
variables interacting linearly. A path integral representation of the likelihood (see e.g.\ \cite{roudi1}) provides the starting point for the Extended Plefka Expansion. The latter 
produces an effectively non-interacting approximation of the original dynamics, more specifically a Gaussian posterior probability that is factorized over hidden nodes but incorporates the hidden-to-observed couplings. In section \ref{sec:pwoPM} we derive equations for the posterior means, which give
the best estimate of the hidden dynamics. In section \ref{sec:pwoPV} we focus then on posterior second moments, i.e.\ hidden-to-hidden correlations, 
hidden responses and auxiliary correlations, in the stationary regime.
In section \ref{quant_tests} we present quantitative tests of the performance of our method, focussing on the predictions for the equal time posterior variance.
We first present the \emph{exact} equations for the posterior statistics, which constitute the baseline for our tests, 
in sections \ref{quant_tests_postvar} and \ref{times}. In section \ref{coup} we describe how the interaction parameters 
are sampled in ours tests; finally in section \ref{sec:results} we show the accuracy of Plefka predictions as a function of the system size, the 
interaction strength and the number of observations.

\section{Extended Plefka Expansion with hidden nodes}
\label{sec:pwoEPEHN}

We consider a generic network where only the dynamics of a subnetwork of nodes is observed while the others are hidden and form what we call the ``bulk''. Such a situation 
could arise because the subnetwork nodes are more precisely characterized from the theoretical point of view, or experimentally more accessible. We assume that we have noise-free data for the trajectories of the observed nodes.   
In biological contexts this is clearly a simplification as data is often available only at discrete time points or 
corrupted by noise. The constraint that only a part of the network can be observed is generic, on the other hand. 
In protein interaction networks, for example, only a few molecular species can typically be tagged biochemically in 
such a way that their concentrations can be tracked with reasonable accuracy \cite{tony,okino}. 

We use the indices $i,j=1,...,N^{\rm b}$ for the hidden or bulk variables and $a,b=1,...,N^{\rm s}$ for the observed or subnetwork nodes of the network, and generally use the superscripts $\rm s$ and $\rm b$ to distinguish variables relating to the observed and hidden sectors, respectively. Assuming that the  hidden and observed variables $x_i$, $x_a$ interact via linear couplings $\lbrace J_{ij} \rbrace$, $\lbrace K_{ia} \rbrace$, their dynamical evolution is described by
\begin{subequations} 
\label{eq:lineqP}
\begin{align}
 \partial_t x_i(t)&=-\lambda x_i(t) +\sum_j J_{ij} x_j(t)+\sum_a K_{ia} x_a(t) +\xi_i(t) \label{eq:lineqP1}\\
 \partial_t x_a(t)&=-\lambda x_a(t) +\sum_b J_{ab} x_b(t)+\sum_j K_{aj} x_j(t) +\xi_a(t) \label{eq:lineqP2}
\end{align}
\end{subequations}
$J_{ij}$ (respectively $J_{ab}$) denotes hidden-to-hidden (respectively observed-to-observed) interactions, while the coupling between observed and hidden variables is contained in $K_{ia}$ and $K_{aj}$.
We have included a self-interaction term with coefficient $\lambda$, which acts as a decay constant and provides the basic timescale of the dynamics. In more compact notation we can write
\begin{subequations} 
\label{eq:lineqPphi}
\begin{align}
 \partial_t x_i(t)&=-\lambda x_i(t) + \phi_i(\bm{x}^{\rm b} (t), \bm{x}^{\rm s}(t)) +\xi_i(t) \label{eq:lineqP1phi}\\
 \partial_t x_a(t)&=-\lambda x_a(t) +\phi_a(\bm{x}^{\rm b} (t), \bm{x}^{\rm s}(t))+\xi_a(t) \label{eq:lineqP2phi}
\end{align}
\end{subequations}
where $\phi_i(\bm{x}^{\rm b} (t), \bm{x}^{\rm s}(t))=\sum_j J_{ij} x_j(t)+\sum_a K_{ia} x_a(t)$ and 
$\phi_a(\bm{x}^{\rm b}(t) , \bm{x}^{\rm s}(t))=\sum_b J_{ab} x_b(t)+\sum_j K_{aj} x_j(t)$ for the linear dynamics \eqref{eq:lineqP} and
$\bm{x}^{\rm b}(t)$ ($\bm{x}^{\rm s}(t)$) denotes the whole set of hidden (observed) values. The main assumption of this setup is that 
the trajectory described by \eqref{eq:lineqP2phi} is known in a finite time window $\lbrace 0, T\rbrace$.
The dynamical noises $\xi_i$, $\xi_a$ are Gaussian white noises with zero mean and diagonal covariances $\Sigma_{i}$, $\Sigma_{a}$
\begin{equation}
\langle \xi_i(t) \xi_j(t')   \rangle= \Sigma_{i}\delta_{ij} \delta(t-t')\qquad \langle \xi_a(t) \xi_b(t')   \rangle= \Sigma_{a} \delta_{ab} \delta(t-t')
\end{equation}
We next summarize the main steps of the derivation of the Extended Plefka Expansion for such a stochastic dynamics. In doing this, we keep 
the generic notation $\phi_i(\bm{x}^{\rm b} (t), \bm{x}^{\rm s}(t))$ and $\phi_a(\bm{x}^{\rm b}(t) , \bm{x}^{\rm s}(t))$ to emphasize that
our treatment does not specifically rely on the particular linear form chosen in \eqref{eq:lineqP}. Indeed the Plefka approach
is applicable to dynamics with any nonlinear drift \cite{bravi}, at the cost of longer expressions in the general case. After discretizing time in elementary time steps $\Delta$, we can write the likelihood -- in our case, the probability of a trajectory of the observed variables -- using
the Martin--Siggia--Rose--Janssen--De Dominicis (MSRJD) path integrals formalism \cite{martin,janssen,dedominicis} as
\begin{eqnarray}
\label{lik0}
\fl P(\bm{x}^{\rm s})&=&\bigg\langle \int D\bm{x}^{\rm b} \prod_{at}\delta\big(x_a(t+\Delta)-x_a(t) -\Delta[-\lambda x_a(t) +\phi_a(\bm{x}^{\rm b}(t), \bm{x}^{\rm s}(t))+\xi_a(t)]\big)\notag\\
\fl&&\prod_{it}\delta\big(x_i(t+\Delta)-x_i(t)- \Delta[-\lambda x_i(t) + \phi_i(\bm{x}^{\rm b} (t), \bm{x}^{\rm s}(t)) +\xi_i(t)]\big)\bigg\rangle_{\bm{\xi}^{\rm s},\,\bm{\xi}^{\rm b} }=\notag\\
\fl&=&\bigg\langle \int D\bm{x}^{\rm b} D\hat{\bm{x}}^{\rm b} D\hat{\bm{x}}^{\rm s}\,\text{exp}\bigg\lbrace\sum_{at}\text{i}\hat{x}_a(t)(x_a(t+\Delta)-x_a(t) -\Delta[-\lambda x_a(t) +\phi_a(\bm{x}^{\rm b} (t), \bm{x}^{\rm s}(t))\notag\\
\fl&+&\xi_a(t)])+ \sum_{it}\text{i}\hat{x}_i(t)(x_i(t+\Delta)-x_i(t)-\Delta[-\lambda x_i(t) + \phi_i(\bm{x}^{\rm b}(t), \bm{x}^{\rm s}(t)) +\xi_i(t)])\bigg\rbrace\bigg\rangle_{\bm{\xi}^{\rm s},\,\bm{\xi}^{\rm b}}\notag=\\
\fl&=&\int D\bm{x}^{\rm b} D\hat{\bm{x}}^{\rm b} D\hat{\bm{x}}^{\rm s}\,\text{e}^{\mathcal{H}}
\end{eqnarray}
where $D\bm{x}^{\rm b} D\hat{\bm{x}}^{\rm b}$ and $D\hat{\bm{x}}^{\rm s}$ are shorthands for the integrations $\prod_{it}dx_i(t)d\hat{x}_i(t)/2\pi$ and $\prod_{at}d\hat{x}_a(t)/2\pi$ 
respectively and the action $\mathcal{H}$ 
\begin{eqnarray}
\label{hcom}
\fl \mathcal{H}&=&\sum_{at}\text{i}\hat{x}_a(t)(x_a(t+\Delta)-x_a(t) -\Delta[-\lambda x_a(t) +\phi_a(\bm{x}^{\rm b} (t), \bm{x}^{\rm s}(t))])-\frac{\Delta}{2}\Sigma_a\hat{x}_a(t)\hat{x}_a(t)\notag\\
\fl &&\sum_{it}\text{i}\hat{x}_i(t)(x_i(t+\Delta)-x_i(t)- \Delta[-\lambda x_i(t) + \phi_i(\bm{x}^{\rm b} (t), \bm{x}^{\rm s}(t))])-\frac{\Delta}{2}\Sigma_i\hat{x}_i(t)\hat{x}_i(t)
\end{eqnarray}
The $\bm{\hat{x}}^{\rm s}(t),\bm{\hat{x}}^{\rm b}(t)$ are auxiliary variables one introduces to represent the $\delta$ function enforcing the dynamics \eqref{eq:lineqP} in terms of an exponential. In the final step of \eqref{lik0} we have
applied a standard Gaussian identity to average over the Gaussian noises, i.e.\ 
\begin{equation}
\langle \text{e}^{\text{i}\Delta \hat{\bm{x}}^{\cdot T}{\bm{\xi}^{\cdot}}} \rangle_{\bm{\xi}^{\cdot}}= \text{e}^{-\Delta\,\bm{\hat{x}}^{\cdot T}\bm{\Sigma}^{\cdot}\bm{\hat{x}}^{\cdot}/2}\qquad 
\cdot= \rm s, \rm b
\end{equation}
where here and below the use of a dot as superscript indicates that the equation is valid for both subnetwork ($\cdot = \rm s$) and bulk ($\cdot = \rm b$). 
We have used the It\^o convention \cite{vankampenito} to discretize the noise and $\xi_i(t)$ above is to be read as the average of the continuous-time noise over the time interval $[t,t+\Delta]$, 
with covariance 
\begin{equation}
\langle {\xi}_i(t){\xi}_i(t') \rangle=\frac{1}{\Delta}\Sigma_{i}\delta_{tt'}\qquad \langle {\xi}_a(t){\xi}_a(t') \rangle=\frac{1}{\Delta}\Sigma_{a}\delta_{tt'}
\end{equation}
Here $\delta_{tt'}/{\Delta}$ is the discrete-time analogue of $\delta(t-t')$.
Note that \eqref{lik0} can be viewed as a partition function that normalizes the posterior distribution of hidden trajectories given the observed trajectory. 
Because of the conditioning on the observations it is not equal to unity, so that one has to be careful not to rely on consequences -- such as the vanishing of all 
moments of $\bm{\hat{x}}^{\rm s}$ and $\bm{\hat{x}}^{\rm b}$ -- that would otherwise follow from this.

The essence of our approach is to treat the interacting terms $\phi_i(\bm{x}^{\rm b}(t), \bm{x}^{\rm s}(t))$ 
and $\phi_a(\bm{x}^{\rm b}(t), \bm{x}^{\rm s}(t))$ within the Extended Plefka Expansion, thus generalizing 
the approach in \cite{bravi} to the presence of observations. The first step is to decide what averages to fix as order parameters for the expansion; we choose the first and second moments of the fluctuating quantities (i.e.\ the ones we integrate over) $\bm{x}^{\rm b}(t)$, $\hat{\bm{x}}^{\rm b}(t)$ and $\hat{\bm{x}}^{\rm s}(t)$. 
Because we assume that the trajectory of the observed variables has been observed, the $\bm{x}^{\rm s}(t)$ are known $\forall t$ and do not need to be estimated.
By analogy with the notation in \cite{bravi} we introduce shorthands for the quantities to be averaged, $\hat{\bm{m}}$,
and for the order parameters they define $\bm{m}= \langle \hat{\bm{m}} \rangle$
\begin{subequations}
\label{shorthandsobs}
\begin{align}
\hat{\bm{m}}^{\rm b} &=\lbrace \bm{x}^{\rm b} , -\text{i}\hat{\bm{x}}^{\rm b} , \bm{x}^{\rm b} \bm{x}^{\rm b} , -\text{i}\hat{\bm{x}}^{\rm b} \bm{x}^{\rm b} , \text{i}\hat{\bm{x}}^{\rm b} \text{i}\hat{\bm{x}}^{\rm b}   \rbrace\\
\hat{\bm{m}}^{\rm s}&=\lbrace -\text{i}\hat{\bm{x}}^{\rm s}, \text{i}\hat{\bm{x}}^{\rm s}\text{i}\hat{\bm{x}}^{\rm s} \rbrace\\
\bm{m}^{\rm b} &=\lbrace \bm{\mu}^{\rm b} ,-\text{i}\hat{\bm{\mu}}^{\rm b} ,\bm{C}^{\rm bb} , \bm{R}^{\rm bb} , \bm{B}^{\rm bb} \rbrace\\
\bm{m}^{\rm s}&=\lbrace -\text{i}\hat{\bm{\mu}}^{\rm s}, \bm{B}^{\rm ss}\rbrace
\end{align}
\end{subequations}
where the components of $\bm{m}^{\rm b}$ and $\bm{m}^{\rm s}$ are explicitly defined as
\begin{subequations}
\label{eq:momentsobs}
\begin{align}
\mu_i(t)&= \langle x_i(t) \rangle\\
\hat{\mu}_i(t)&= \langle \hat{x}_i(t) \rangle\\
\hat{\mu}_a(t)&= \langle \hat{x}_a(t) \rangle\\
C_{i}(t, t')&=  \langle x_i(t)x_i(t')\rangle\\
R_{i}(t', t)&=  -\text{i}\langle\hat{x}_i(t)x_i(t') \rangle\\
B_{i}(t, t')&= - \langle\hat{x}_i(t)\hat{x}_i(t')\rangle\\
B_{a}(t, t')&= - \langle\hat{x}_a(t)\hat{x}_a(t')\rangle
\end{align}
\end{subequations}
These averages are the posterior moments over physical and auxiliary dynamical trajectories. 
Note that we keep only the diagonal second moments:
this is crucial in order to obtain an effective non-interacting description from the Plefka expansion. This choice implies that different sites,
including the observed $a$ and hidden ones $i$, are effectively decoupled at the level of correlations and responses, with the contribution from interactions being 
captured in appropriately defined single site fields. As a result, this implementation of the Plefka expansion gives a 
\emph{mean-field type approximation}. The starting point of the expansion (see e.g.\ \cite{bravi}) is in fact to augment the original partition function 
$P(\bm{x}^{\rm s})$ with field terms -- denoted in our context $\bm{h}_{\ap}^{\rm b}$ and $\bm{h}_{\ap}^{\rm s}$ -- conjugate to the chosen observables, and then consider the Legendre transform of the
corresponding free energy
\begin{equation}
\label{eq:gammaxiobs}
G_{\ap}(\bm{m}^{\rm s},\bm{m}^{\rm b} ,\bm{x}^{\rm s}) =
\ln{ \int D\bm{x}^{\rm b}  D\hat{\bm{x}}^{\rm b}  D\hat{\bm{x}}^{\rm s}\, \text{e}^{\Xi_{\ap}}}
\end{equation}
where
\begin{equation}
\label{fullxi0}
 \Xi_{\ap}=\mathcal{H}_{\ap}+\bm{h}_{\ap}^{\rm b} (\hat{\bm{m}}^{\rm b} -\bm{m}^{\rm b} )+ \bm{h}_{\ap}^{\rm s}(\hat{\bm{m}}^{\rm s}-\bm{m}^{\rm s})
\end{equation}
The key of the Plefka expansion is the introduction of the
parameter $\ap$ here, which scales the interacting parts of the Hamiltonian. For $\ap=0$ one then has a non-interacting
system and the Plefka expansion is a perturbation expansion of $G_{\ap}$ around this point, where one sets $\ap=1$ at the end to 
recover the full  Hamiltonian $\mathcal{H}\equiv \mathcal{H}_1$.

In our model, $\Xi_{\ap}$ is given explicitly by
\begin{eqnarray}
\fl \Xi_{\ap}&&=\sum_{it} \text{i}\hat{x}_i(t)\big(x_i(t+\Delta)-x_i(t)+\Delta\lambda x_i(t)-\ap\Delta\phi_i(\bm{x}^{\rm b} (t),\bm{x}^{\rm s}(t))\big)+ \notag\\
\fl &&+\sum_{at} \text{i}\hat{x}_a(t)\big(x_a(t+\Delta)-x_a(t)+\Delta\lambda x_a(t)-\ap\Delta\phi_a(\bm{x}^{\rm b} (t), \bm{x}^{\rm s}(t))\big)+ \notag\\
\fl &&+\Delta\sum_{it}\psi_{i\ap}(t)\big(x_i(t)-\mu_i(t)\big)-\Delta\sum_{it}l_{i\ap}(t)\big(\text{i}\hat{x}_i(t)-\text{i}\hat{\mu}_i(t)\big)-\Delta\sum_{at}l_{a\ap}(t)\big(\text{i}\hat{x}_a(t)-\text{i}\hat{\mu}_a(t)\big)+\notag\\
\fl &&+\Delta^2\sum_{itt'}\hat{C}_{i\ap}(t,t')\big(x_i(t)x_i(t')-C_i(t,t')\big)+\Delta^2\sum_{itt'}\hat{R}_{i\ap}(t,t')\big(-\text{i}\hat{x}_i(t)x_i(t')-R_i(t',t)\big)+\notag\\
\fl &&+\frac{\Delta^2}{2}\sum_{itt'}\hat{B}_{i\ap}(t,t')\big(-\hat{x}_i(t)\hat{x}_i(t')-  B_{i}(t,t')\big)+\frac{\Delta^2}{2}\sum_{att'}\hat{B}_{a\ap}(t,t')\big(-\hat{x}_a(t)\hat{x}_a(t')-  B_{a}(t,t')\big)+\notag\\
\fl &&-\frac{\Delta}{2}\sum_{it}\Sigma_{i}\hat{x}_i(t)\hat{x}_i(t)-\frac{\Delta}{2}\sum_{at}\Sigma_{a}\hat{x}_a(t)\hat{x}_a(t)
\label{eq:xialphaobs}
\end{eqnarray}
The first two lines and the last give $\mathcal{H}_{\ap}$, the Hamiltonian for the interacting dynamics $\eqref{eq:lineqP}$ but with the interaction 
terms $\phi_i(\bm{x}^{\rm b}(t), \bm{x}^{\rm s}(t))$ and $\phi_a(\bm{x}^{\rm b}(t), \bm{x}^{\rm s}(t))$ scaled by $\ap$. By definition of the Legendre 
transform, one can derive the conjugate fields as
\be
\bm{h}_{\ap}^{\cdot}=-\frac{1}{\Delta^n}\frac{\partial G_{\ap}}{\partial\bm{m}^{\cdot}} \qquad \cdot=\rm s, \rm b
\label{eq:hobs_generic}
\ee
where as in (\ref{eq:xialphaobs}) we choose $n=1$ for linear fields and $n=2$ for quadratic ones to obtain well-defined values in the continuous time limit $\Delta\to 0$. The fields have components 
\begin{subequations}
\label{shorthandsfieldsobs}
\begin{align}
\bm{h}_{\ap}^{\rm b} &=\lbrace \bm{\Psi}_{\ap}^{\rm b} , \bm{l}_{\ap}^{\rm b} , \hat{\bm{C}}_{\ap}^{\rm bb} , \hat{\bm{R}}_{\ap}^{\rm bb} , \hat{\bm{B}}_{\ap}^{\rm bb}  \rbrace\\
\bm{h}_{\ap}^{\rm s}&=\lbrace \bm{l}_{\ap}^{\rm s}, \hat{\bm{B}}_{\ap}^{\rm ss} \rbrace
\end{align}
\end{subequations}
that are given explicitly by
\begin{subequations}
\label{fields}
\begin{align}
\psi_{i\ap}(t)&=-\frac{1}{\Delta}\frac{\partial G_{\ap}}{\partial \mu_i(t)}\\
-\text{i}l_{i\ap}(t)&=-\frac{1}{\Delta}\frac{\partial G_{\ap}}{\partial (\hat{\mu}_i(t))}\\
-\text{i}l_{a\ap}(t)&=-\frac{1}{\Delta}\frac{\partial G_{\ap}}{\partial (\hat{\mu}_a(t))}\\
\hat{R}_{i\ap}(t,t') &= -\frac{1}{\Delta^2}\frac{\partial G_{\ap}}{\partial R_i(t',t)}\\
\hat{C}_{i\ap}(t,t') &= -\frac{1}{\Delta^2}\frac{\partial G_{\ap}}{\partial C_i(t,t')}\\
\hat{B}_{i\ap}(t,t') &= -\frac{1}{\Delta^2}\frac{\partial G_{\ap}}{\partial B_{i}(t,t')}\\
\hat{B}_{a\ap}(t,t') &= -\frac{1}{\Delta^2}\frac{\partial G_{\ap}}{\partial B_{a}(t,t')}
\end{align}
\end{subequations}
The original dynamics has no biasing fields so the condition that defines the physical values of the order parameters is simply
\be
\bm{h}_{\ap}^{\cdot}=0\qquad \cdot = \rm s, \rm b
\ee
If one now Taylor expands to second order in $\alpha$ as $G_\alpha=G^0+\alpha G^1+(\alpha^2/2)G^2$, and similarly for $\bm{h}_\alpha$, then one sees that the physical order parameter values are
obtained in the {\em non-interacting} ($\alpha=0$) theory by applying effective fields given by
\be 
\label{eq:zerofields}
\bm{h}^{\rm eff\, \cdot} = -\ap \bm{h}^{1\,\cdot} -\frac{\ap^2}{2}\bm{h}^{2\,\cdot} \qquad \cdot=\rm s, \rm b
\ee 
One has from \eqref{fullxi0}
\be
\label{eq:der}
\frac{d\Xi_{\ap}}{d \ap}= \frac{\partial \mathcal{H}_{\ap}}{\partial \ap}+ \frac{\partial \bm{h}_{\ap}^{\rm b}}{\partial \ap} (\hat{\bm{m}}^{\rm b} -\bm{m}^{\rm b} )
+\frac{\partial \bm{h}_{\ap}^{\rm s}}{\partial \ap}(\hat{\bm{m}}^{\rm s}-\bm{m}^{\rm s})
\ee
where one defines $\partial_{\alpha} \bm{h}_{\ap}^{\rm \cdot}= \bm{h}^{1 \rm \cdot}$, with $\cdot = \rm s,b$ as before. 
Inserting this into the general expression for $G^1\equiv\partial G_{\alpha}/\partial \alpha\vert_{\alpha=0}$ (see \cite{bravi}) gives
\begin{eqnarray}
\label{eq:g1g}
\fl G^1&=&\bigg\langle \frac{d\Xi_{\alpha}}{d \alpha} \bigg\rangle_{0} =\bigg\langle-\Delta\bigg[\sum_{at}\text{i}\hat{x}_a(t)\phi_{a}(\bm{x}^{\rm b} (t),\bm{x}^{\rm s}(t)) +  \sum_{it}\text{i}\hat{x}_i(t)
\phi_{i}(\bm{x}^{\rm b} (t),\bm{x}^{\rm s}(t))\bigg]\bigg\rangle_0
\end{eqnarray}
where the average of the factors multiplying $\bm{h}^{1 \rm \cdot}$ with $\cdot = \rm s,b$ vanishes by definition \eqref{eq:momentsobs}.
If we specialize again to the linear dynamics $\eqref{eq:lineqP}$, $G^1$ becomes 
\begin{eqnarray}
\label{eq:g1}
\fl &&G^1=\\
\fl&&-\Delta \bigg[\sum_{at} \text{i}\hat{\mu}_a(t)\bigg(\sum_b J_{ab} x_b(t)+\sum_j K_{aj} \mu_j(t)\bigg)  +  
\sum_{it}\text{i}\hat{\mu}_i(t)\bigg(\sum_j J_{ij}\mu_j(t)+\sum_a K_{ia} x_a(t) \bigg)\bigg]\notag
\end{eqnarray}
For $G^2$ we restrict ourselves directly to the linear case to avoid lengthier expressions; these are given for reasonably general 
nonlinear drift in \ref{appendix:a}. We then need
\begin{equation}
\label{eq:da2}
\fl \delta\bigg(\frac{d \Xi_{\ap}}{d \ap} \bigg)=\frac{d \Xi_{\ap}}{d \ap}-\bigg\langle \frac{d \Xi_{\ap}}{d \ap} \bigg\rangle_{0}=
-\Delta \bigg [\sum_{ajt}\text{i}\delta\hat{x}_a(t)K_{aj}\delta x_j(t)+\sum_{ijt}\text{i}\delta\hat{x}_i(t)J_{ij}\delta x_j(t)\bigg]
\end{equation}
In writing this we have used the expression for the fields $\bm{h}^{1 \rm \cdot}$ for the linear case. Assuming 
there are no self-interactions ($J_{aa}=J_{ii}=0$), these read 
\begin{subequations}
\begin{align}
\psi_i^1(t)&= \sum_a \text{i}\hat{\mu}_a(t)K_{ai} + \sum_{j} \text{i}\hat{\mu}_{j}(t)J_{ji}\\
l_i^1(t)&=-\sum_jJ_{ij}\mu_j(t)-\sum_a K_{ia}x_a(t)\\
l_a^1(t)&=-\sum_jK_{aj}\mu_j(t)-\sum_b J_{ab}x_b(t)\\
\hat{R}_i^1(t,t') &= \hat{C}_i^1(t,t')=\hat{B}_i^1(t,t')=\hat{B}_a^1(t,t')=0
\end{align}
\label{eq:fieldsl}
\end{subequations}
as can be derived by applying \eqref{eq:hobs_generic} to $G^1$ \eqref{eq:g1}.
Inserting \eqref{eq:fieldsl} into \eqref{eq:der} and simplifying, one finds \eqref{eq:da2}, where all terms containing 
non-fluctuating quantities (i.e.\ observations) have vanished.
$G^2\equiv\partial^2 G_{\alpha}/\partial \alpha^2\vert_{\alpha=0}$ can now be calculated \cite{bravi} as
\begin{eqnarray}
\label{eq:g2}
\fl G^2&=&\bigg \langle \bigg(\delta\bigg(\frac{d \Xi_{\ap}}{d \ap} \bigg)\bigg)^2 \bigg \rangle_{0} =\\
\fl &=&\Delta^2 \bigg [\sum_{ajtt'}K_{aj}^2\delta \hc_a(t,t')\delta C_j(t, t')+\sum_{ijtt'}J_{ij}^2 \delta \hc_i(t,t')\delta C_j(t, t') +\sum_{ijtt'}J_{ij}J_{ji}\delta R_i(t,t')\delta R_j(t',t)\bigg]\notag
\end{eqnarray}
where $\delta C_i(t, t')$, $\delta \hc_{a/i}(t,t')$ and $\delta R_i(t,t')$ denote the connected correlators and responses, e.g. $\delta C_i(t, t') =  C_i(t, t')- \mu_i(t)\mu_i(t')$. 
All averages of products decouple into averages of variables at different nodes as
$J_{aa}=J_{ii}=0$. In addition, when the square is calculated, some of the contractions among terms are zero because of the factorization among different nodes in the effective dynamics at $\alpha=0$. The last term, which contains
anti-causal responses, must be retained at this level as it contributes to the derivatives \eqref{fields} defining the fields.
By taking the derivatives of $G^2$ we can find the second order contribution to the effective fields \eqref{eq:zerofields}; combining with the first order terms above gives 
(at $\ap=1$)
\begin{subequations}
\label{eq:eff2m}
\begin{align}
\begin{split}
 l^{\rm eff}_i(t)=&\sum_jJ_{ij}\mu_j(t)+\sum_a K_{ia}x_a(t)+\int_0^{T} dt'\sum_{j}\text{i}\hat{\mu}_i(t') J_{ij}^2\delta C_j(t, t') \\
 &- \int_0^{t}dt'\sum_{j}J_{ij}J_{ji}\mu_i(t')\delta R_j(t,t')
\end{split}\\
\begin{split}
 l^{\rm eff}_a(t)=&\sum_jK_{aj}\mu_j(t)+\sum_b J_{ab}x_b(t)+\int_0^{T} dt' \sum_{a}\text{i}\hat{\mu}_a(t') K_{aj}^2\delta C_j(t, t')
\end{split}\\
\begin{split}
 \psi^{\rm eff}_i(t)=& -\sum_a \text{i}\hat{\mu}_a(t)K_{ai} - \sum_{j} \text{i}\hat{\mu}_{j}(t)J_{ji}+
 \int_{t}^{T}dt'\sum_{j}J_{ij}J_{ji}\delta R_j(t',t) \text{i}\hat{\mu}_i(t')\\
 &-\int_0^{T} dt' \bigg( \sum_{a}K_{ai}^2\delta \hc_a(t,t') \mu_i(t')+ \sum_{j}J_{ji}^2 \delta \hc_j(t,t') \mu_i(t')\bigg) 
\end{split}\\
\begin{split}
 \hat{R}^{\rm eff}_i(t',t)& =\sum_{j}J_{ij}J_{ji}\delta R_j(t',t)\label{eq:effres}
\end{split}\\
\begin{split}
\hat{C}^{\rm eff}_i(t,t')&=\sum_{a}K_{ai}^2 \delta \hc_a(t,t') + \sum_{j}J_{ji}^2 \delta \hc_j(t,t')
\end{split}\\
\begin{split}
\hat{B}^{\rm eff}_i(t,t')&=\sum_{j}J_{ij}^2 \delta C_j(t,t')\label{eq:effcorr}
\end{split}\\
\hat{B}^{\rm eff}_a(t,t')&=\sum_{j}K_{aj}^2 \delta C_j(t,t')
\end{align}
\end{subequations} 
These can be substituted into $\Xi_0$ (i.e.\ \eqref{eq:xialphaobs} at $\alpha=0$) to give the Plefka approximation for our partition function
\be
\label{lik}
P^{\rm eff}(\bm{x}^{\rm s})=\int D\bm{x}^{\rm b}  D\hat{\bm{x}}^{\rm b}  D\hat{\bm{x}}^{\rm s}\,\text{e}^{\mathcal{H}^{\rm eff}}
\ee
with
\begin{eqnarray}
\label{heff}
\fl \mathcal{H}^{\rm eff} &&= \sum_{at}\text{i}\hat{x}_a(t)(x_a(t+\Delta)-x_a(t) +\Delta\lambda x_a(t) -\Delta l^{\rm eff}_a(t))-\frac{\Delta^2}{2}\sum_{att'}(\hat{B}^{\rm eff}_a(t,t')
+\frac{\Sigma_a}{\Delta}\delta_{tt'})\hat{x}_a(t)\hat{x}_a(t')\notag\\
\fl &&+\sum_{it}\text{i}\hat{x}_i(t)(x_i(t+\Delta)-x_i(t) + \Delta\lambda x_i(t) - \Delta l^{\rm eff}_i(t) -\Delta^2\sum_{t'}\hat{R}^{\rm eff}_i(t,t')x_i(t'))\\
\fl &&-\frac{\Delta^2}{2}\sum_{itt'}(\hat{B}^{\rm eff}_i(t,t')+\frac{\Sigma_i}{\Delta}\delta_{tt'})\hat{x}_i(t)\hat{x}_i(t')+\Delta\sum_{it}\psi^{\rm eff}_i(t)x_i(t)+\frac{\Delta^2}{2}
\sum_{itt'}\hat{C}^{\rm eff}_i(t,t')x_i(t)x_i(t')\notag
\end{eqnarray}
The approximating action $\mathcal{H}^{\rm eff}$ is factorized over sites, as anticipated above, thus
the dynamics resulting from the action of the effective fields is non-interacting.
As \eqref{heff} is quadratic, it gives a Gaussian weight for the trajectories at each site.\\
Expression \eqref{lik} then shows how the  effective fields feature in the approximate posterior (conditional on the observations) dynamics. 
For the hidden nodes $x_i(t)$, similarly to \cite{bravi}, one sees that $l^{\rm eff}_i(t)$ is an effective drift; an 
additional coloured Gaussian noise also acts on $x_i(t)$, with covariance $\hat{B}^{\rm eff}_i(t,t')$. Finally
$\hat{R}^{\rm eff}_i(t,t')$, as given by \eqref{eq:effres}, is a memory kernel with a simple intuitive interpretation: a fluctuation $\delta x_i$ at time $t'$ acts via $J_{ji}$ as an effective 
field on $x_j$; at time $t$ this produces a response in $x_j$ modulated by $R_j(t,t')$, which then acts back on
$x_i$ via $J_{ij}$. The presence of this term might seem to disagree with the dynamical mean-field theory of asymmetric spin networks obtained by
Kappen and Spanjers \cite{kappen} and M\'ezard and Sakellariou \cite{inference3}. However, as shown by Bachschmid-Romano et al. \cite{romanobattistin}, this is only an apparent contradiction, and the Extended Plefka Expansion does give back those known results. This is because in the limit of large $N^{\rm b}$
the memory with coefficient $J_{ij}J_{ji}$ can be neglected for asymmetric couplings. On the other hand
a $J_{ij}^2$ term, which does appear in Refs.~\cite{kappen,inference3} is left over from the coloured noise \eqref{eq:effcorr}. In addition, we note that also the observed dynamics acquires effective fields, with $l^{\rm eff}_a(t)$ a linear drift and 
$\hat{B}^{\rm eff}_a(t,t')$ the correlator of a Gaussian coloured noise on $x_a(t)$.
Finally, the fields $\psi^{\rm eff}_i(t)$ and $\hat{C}^{\rm eff}_i(t,t')$, which would be zero without observations \cite{bravi}, 
effectively constrain the hidden dynamics to be consistent with the observed trajectories, as we will explicitly show in the next section.

As a result of including first and second moments (i.e.\ means, responses and correlations), the statistics of hidden (physical and auxiliary) and observed (auxiliary) 
trajectories is Gaussian within our 
approximation. The exact posterior for our linear dynamics is also Gaussian, so in this case the mean-field approximation consists in assuming
site-diagonal second moments. The posterior means of the hidden variables can be regarded as the ``best" estimate 
-- in the mean-square sense\footnote{The ``best" or ``optimal" estimate in the mean-square sense 
is the prediction that minimizes the mean square error between
the actual and estimated data (the hidden trajectories $x_i(t)$ in this case). 
Importantly, the Kalman filter \cite{kalmanor} is an algorithm designed to compute this optimal predictor \emph{exactly}.} --
of the hidden dynamics, while the equal time posterior variance quantifies the degree of 
uncertainty for those inferred values and hence the inference error. 
Both the means and variances can be read off from the integrand in \eqref{lik}, which is proportional to the approximate Gaussian posterior over hidden trajectories conditioned on observations.
The effective fields that couple to linear observables  ($\psi^{\rm eff}_i(t),l^{\rm eff}_a(t),l^{\rm eff}_i(t)$) determine the posterior means; the ones that correspond to quadratic observables ($\hat{C}^{\rm eff}_i(t,t'),\hat{R}^{\rm eff}_i(t,t'),\hat{B}^{\rm eff}_i(t,t'),\hat{B}^{\rm eff}_a(t,t')$) characterize the posterior second moments, i.e.\ posterior responses 
and correlations.

\subsection{Posterior Means}
\label{sec:pwoPM}

For a Gaussian distribution in general the mean is the point where the probability density is stationary. To find the posterior mean we therefore just have to set 
the derivatives of the effective action to zero
\begin{equation}
\frac{\partial\mathcal{H}^{\rm eff}}{\partial \hat{x}_i(t)}=0,
\qquad
 \frac{\partial \mathcal{H}^{\rm eff}}{\partial x_i(t)}=0
\label{eq:stationarity}
\end{equation}
One finds from these conditions (already in the continuous limit $\Delta \rightarrow 0$)
\begin{eqnarray}
\label{meandyn}
\fl \partial_t \mu_i(t)& = & -\lambda \mu_i(t) + \lhe +\int_{0}^t dt'\rhe\mu_i(t') -\int_{0}^{T} dt'\bigg(\hat{B}^{\rm eff}_i(t,t')+\Sigma_i\delta(t-t')\bigg)\text{i}\hat{\mu}_i(t')\notag \\
\fl &=& -\lambda \mu_i(t) + \sum_j J_{ij}\mu_j(t)+\sum_a K_{ia}x_a(t)-\Sigma_i\text{i}\hat{\mu}_i(t)
\end{eqnarray}
which is the \emph{exact} dynamics for the bulk plus an additional term with $\hat{\mu}_i(t)$ from conditioning on observations, 
which can also be viewed as a nonzero mean of the noise 
caused by conditioning. Another interpretation of
$\hat{\mu}_i(t)$ is as a back-propagating error, as can be seen from its dynamics. The latter follows from the second part 
of (\ref{eq:stationarity}) as 
\begin{eqnarray}
\label{meandynhat}
\fl \partial_t(\text{i}\hat{\mu}_i(t))&=&\lambda(\text{i}\hat{\mu}_i(t))+\psi_i^{\rm eff}(t)-\int_t^{T}dt'\hat{R}^{\rm eff}_i(t',t)(\text{i}\hat{\mu}_i(t'))+
\int_0^{T}dt'\hat{C}^{\rm eff}_i(t,t')\mu_i(t')\notag\\
\fl &=&\lambda(\text{i}\hat{\mu}_i(t))-\sum_a \text{i}\hat{\mu}_a(t)K_{ai} - \sum_{j} \text{i}\hat{\mu}_{j}(t)J_{ji}
\end{eqnarray}
We can now more clearly understand the roles played by the effective fields in the auxiliary dynamics, with $\psi^{\rm eff}_i(t)$ being an effective drift 
and $\hat{C}^{\rm eff}_i(t,t')$ the covariance of an additional coloured noise. The integration over $T > t'>t$ indicates that
$\hat{R}^{\rm eff}_i(t',t)$ can also be interpreted as a memory kernel in this context, but it is a memory 
``from the future'' in the sense that it provides the weight with which the future values 
of $\hat{\mu}_i(t)$ affect the present ones. The dynamics itself of $\hat{\mu}_i(t)$ propagates its values backwards
from the final time $T$. This dynamical evolution depends on the $\hat{\mu}_a(t)$, which in turn \emph{does depend} on the dynamics of observations:
in this way, observations make auxiliary variables non-zero. In fact one has from the analogue for observations of conditions \eqref{eq:stationarity},
i.e.\ $\partial\mathcal{H}^{\rm eff}/\partial \hat{x}_a(t)=0$,
\begin{eqnarray}
\label{meandynhata}
\fl \partial_t x_a(t)&=& - \lambda x_a(t)+ l^{\rm eff}_a(t) -\int_{0}^{T} dt'\bigg(\hat{B}^{\rm eff}_a(t,t')+\Sigma_a\delta(t-t')\bigg)\text{i}\hat{\mu}_a(t')\notag\\
\fl &=&- \lambda x_a(t)+\sum_jK_{aj}\mu_j(t)+\sum_b J_{ab}x_b(t)-\Sigma_a\text{i}\hat{\mu}_a(t)
\end{eqnarray}
To better grasp how the conditioning on observations enters the dynamics as a backward propagation, we shall briefly go back to the discrete time version
of equations \eqref{meandyn}, \eqref{meandynhat}, \eqref{meandynhata}
\be
\label{meandyn_discrete}
\fl \mu_i(t+\Delta)-\mu_i(t)=-\lambda \mu_i(t) + \sum_j J_{ij}\mu_j(t)+\sum_a K_{ia}x_a(t)-\Sigma_i\text{i}\hat{\mu}_i(t)
\ee
\be
\label{meandynhat_discrete}
\fl \text{i}\hat{\mu}_i(t)-\text{i}\hat{\mu}_i(t-\Delta)=
\lambda(\text{i}\hat{\mu}_i(t))-\sum_a \text{i}\hat{\mu}_a(t)K_{ai} - \sum_{j} \text{i}\hat{\mu}_{j}(t)J_{ji}
\ee
\be
\label{meandynhata_discrete}
\fl x_a(t+\Delta)- x_a(t)=- \lambda x_a(t)+\sum_jK_{aj}\mu_j(t)+\sum_b J_{ab}x_b(t)-\Sigma_a\text{i}\hat{\mu}_a(t)
\ee
where $\Delta$ is the unit time step and the time index $t$ runs from 
$\Delta,...,T$ for $\lbrace \mu_i(t) \rbrace$ while $t=0,...,T-\Delta$ for 
$\lbrace \hat{\mu}_i(t) \rbrace$ and $\lbrace \hat{\mu}_a(t) \rbrace$. Note that \emph{by construction} of the path integral representation
of the dynamics the auxiliary variables are only defined up to time $T-\Delta$ and 
accordingly \eqref{meandynhat_discrete} at $t=T$ has to be read with all 
$\hat{\mu}_i(T)$ and $\hat{\mu}_a(T)$ terms absent
(for more details on the formalism see e.g.\ the general review on path integral 
methods \cite{PathMethods}).
This then gives $\hat{\mu}_i(T-\Delta)=0$, hence the ``initial'' condition is at
the end and this sets the direction of time integration. 
The analogous ``initial'' condition $\hat{\mu}_a(T-\Delta)$ is determined by the observed values at the end of the trajectory $x_a(T)$, $x_a(T-\Delta)$, 
as is clear from \eqref{meandynhata_discrete}.

Equation \eqref{meandynhat_discrete} calculated at $t=T$ and $t=T-\Delta$
shows $\hat{\mu}_i(T-\Delta)$ vanishes but $\hat{\mu}_i(T-2\Delta)$ assumes a value different from
zero because of $\hat{\mu}_a(T-\Delta)$; in particular $\hat{\mu}_i(t-2\Delta)=\sum_a \hat{\mu}_a(t-\Delta)K_{ai}$. This closer inspection
time step by time step immediately reveals that the fixed values of observations introduce a non-zero correction term $\hat{\mu}_i(t)$ in the 
average bulk dynamics \eqref{meandyn_discrete} that evolves backward in time (i.e.\ the ``initial'' values, which are zero, are at $t=T-\Delta$).\\
Together, the above equations for the posterior means $\eqref{meandyn}$, $\eqref{meandynhat}$, $\eqref{meandynhata}$ constitute a forward-backward
propagation, with the $\hat\mu_i$ responsible for the backward portion, i.e.\ for the flow of information from the future. The 
forward-backward structure is consistent with the general theory of conditional stochastic processes, for which the estimation of posterior distributions requires information to propagate both in the
forward and backward direction \cite{stratonovich}.

To recapitulate, the averages of auxiliary variables do not vanish identically as in 
the Extended Plefka Expansion without observations \cite{bravi}, where the mean (i.e.\ stationary) path of a linear dynamics 
would be given by a noiseless relaxation. The observed trajectories effectively ``bias'' the path ensemble in such a way that the mean hidden conditional
dynamics follows a path where auxiliary variables are \emph{non-zero}: 
these can therefore be seen as implementing the constraints given by the observations. 
This is compatible with the fact that the path integral \eqref{lik0} is not expected to be equal to one since it has the meaning of
a data likelihood instead of the normalization of a probability distribution.

The expressions for the means $\eqref{meandyn}$, $\eqref{meandynhat}$, $\eqref{meandynhata}$ are expected to be exact for this linear dynamics 
we study as the Plefka approximation consists of neglecting the off-diagonal terms \emph{only} in the second moments, e.g.\ cross-correlations, 
which do not appear in the equations for the means. To verify this, one notes that the
exact action \eqref{hcom} --
defining a Gaussian path integral in an enlarged space containing auxiliary variables and containing
all, diagonal and off-diagonal, correlations and responses -- is {\em quadratic}. The exact means can therefore be obtained as the saddle point of this action, and this leads to 
equations identical to \eqref{meandyn}, \eqref{meandynhat}, \eqref{meandynhata}.

In addition, as these equations give the exact posterior means 
defined by the Gaussian path integral, their solution must coincide with the exact solution
computable from the Kalman filter recursion (see \cite{thesis} for a direct comparison of the equations). 
We remark, however, that the Plefka approach is more widely applicable than just to the linear case and provides 
an approximate, self-consistent way of estimating the time evolution of the posterior means when they are not exactly tractable, i.e.\ for
generic nonlinear drifts.

\subsection{Posterior Variance}
\label{sec:pwoPV}
The inverse covariance of the approximating Gaussian distribution can be read off from the path integral representation of the likelihood \eqref{lik}. In particular it consists of
two distinct blocks, as the distribution is factorized w.r.t.\ the subnetwork ${\rm s}$ and bulk ${\rm b}$ indices. These blocks can be written symbolically as 
\[\bm{C}_{i\,\text{gen}}^{-1}(t,t')=\begin{pmatrix}
-\hat{C}^{\rm eff}_i(t,t')&(\partial_{t'}+\lambda)\delta(t-t')-\hat{R}^{\rm eff}_i(t',t)\\
(\partial_{t}+\lambda)\delta(t-t')-\hat{R}^{\rm eff}_i(t,t')&-\hat{B}^{\rm eff}_i(t,t')-\Sigma_i\delta(t-t')\end{pmatrix}\]
\[C_{a\,\text{gen}}^{-1}(t,t')=-\hat{B}^{\rm eff}_a(t,t')-\Sigma_a\delta(t-t')\]
The ``\text{gen}'' subscript indicates a ``generalized'' inverse covariance including the auxiliary variables $\hat{x}_i(t)$, $\hat{x}_a(t)$,
i.e.\ 
\[\bm{C}_{i\,\text{gen}}(t,t') =\Bigg\langle\begin{pmatrix}
  \delta x_i(t) \\
   -\text{i}\delta\hat{x}_i(t)\\
\end{pmatrix}\begin{pmatrix}   \delta x_i(t') &-\text{i}\delta\hat{x}_i(t')
\end{pmatrix}\Bigg\rangle\]
while $C_{a\,\text{gen}}(t,t')=-\langle \delta \hat{x}_a(t) \delta \hat{x}_a(t')\rangle$ 
refers {\em only} to the auxiliary variables $\hat{x}_a(t)$ as the $x_a(t)$ are fixed by the observations (here 
$\delta x_i(t) $ indicates the deviation from the mean, i.e.\ $\delta x_i(t)= x_i(t)-\mu_i(t)$ and similarly for 
$\delta\hat{x}_i(t)$ and $\delta\hat{x}_a(t)$). In principle all of the second order functions are connected but in what follows we drop the $\delta$s for the sake of brevity.
The covariance itself then has the same block structure
\[\bm{C}_{i\,\text{gen}}(t,t')=\begin{pmatrix}
C_i(t,t')&R_i(t,t')\\
R_i(t',t)& B_i(t,t')\end{pmatrix}\]
\[C_{a\,\text{gen}}(t,t')=B_a(t,t')\]
Note that here and below we write directly the continuous time equations that are obtained from the discrete time formalism as $\Delta\to 0$ 
(see \cite{thesis} for more details).
To find the equations for the (generalized) covariances, one just takes the identity 
$\int d\tau\,\textbf{C}_{i\,\text{gen}}(t,\tau)\textbf{C}_{i\,\text{gen}}^{-1}(\tau,t')=\mathbb{1}\delta(t-t')$ block by block to get 
\small
\begin{subequations}
\label{integeq}
\begin{align}
&\int_{-\infty}^{+\infty}d\tau \bigg[-C_i(t,\tau)\hat{C}^{\rm eff}_i(\tau,t')+ R_i(t,\tau)\bigg((\partial_{\tau}+\lambda)\delta(\tau-t')-\hat{R}^{\rm eff}_i(\tau,t')\bigg)\bigg]=\delta (t-t')\label{intc1}\\
&\int_{-\infty}^{+\infty}d\tau \bigg[C_i(t,\tau)\bigg((\partial_{t'}+\lambda)\delta(t'-\tau)-\hat{R}^{\rm eff}_i(t',\tau)\bigg) -R_i(t,\tau)\bigg(\hat{B}^{\rm eff}_i(\tau,t')+
\Sigma_i \delta(\tau-t')\bigg)\bigg]= 0\label{intr1}\\
& \int_{-\infty}^{+\infty}d\tau \bigg[R_i(\tau,t)\bigg((\partial_{t'}+\lambda)\delta(t'-\tau)-\hat{R}^{\rm eff}_i(t',\tau)\bigg)-\hc_{i}(t,\tau)\bigg(\hat{B}^{\rm eff}_i(\tau,t')+
\Sigma_i\delta(\tau-t') \bigg) \bigg]= \delta (t-t')\notag\label{intb1}\\
\end{align}
\end{subequations}
\normalsize
and similarly for the auxiliary variables
\begin{equation}
\label{eqca}
\int_{-\infty}^{+\infty} d\tau\,C_{a\,\text{gen}}(t,\tau)C_{a\,\text{gen}}^{-1}(\tau,t')=- \int_{-\infty}^{+\infty}d\tau \,\hc_a(t,\tau)\bigg[\hat{B}^{\rm eff}_a(\tau,t')+\Sigma_a\delta (\tau-t')\bigg]=\delta (t-t')
\end{equation} 
If we now substitute the expressions for $\hat{R}^{\rm eff}_i(t',t)$, $\hat{C}^{\rm eff}_i(t,t')$, $\hat{B}^{\rm eff}_i(t,t')$ and $\hat{B}^{\rm eff}_a(t,t')$ from \eqref{eq:eff2m} we get a closed system of integral equations. To simplify these, we consider long times, where a stationary regime is reached so that all two-time functions become 
Time Translation Invariant (TTI). In an inference problem, stationarity implies also that the quality of the prediction is the same at all times.
At stationarity, the integrals in the above equations \eqref{integeq} and \eqref{eqca} become convolutions and hence, if we go to double-sided Laplace transform, simple products. After some 
simplification the Laplace-transformed equations yield a system of four coupled equations 
for $\tilde{C}_i(z)$, $\tilde{R}_i(z)$, $\tilde{\hc}_{i}(z)$, $\tilde{\hc}_a(z)$
\begin{subequations}
\label{eqs0}
\begin{align}
&-\tilde{C}_i(z)\bigg(\sum_{a}K_{ai}^2 \tilde{\hc}_{a}(z)+\sum_{j}J_{ji}^2 \tilde{\hc}_{j}(z)\bigg)+\tilde{R}_i(z)\bigg(z+\lambda-\sum_jJ_{ij}J_{ji}\tilde{R}_j(z)\bigg)=1\label{uffa1}\\
&\tilde{C}_i(z)\bigg(-z+\lambda-\sum_jJ_{ij}J_{ji}\tilde{R}_j(-z)\bigg)-\tilde{R}_i(z)\bigg(\sum_{j}J_{ij}^2\tilde{C}_j(z)+\Sigma_i\bigg)=0\label{uffa2}\\
&\tilde{\hc}_{i}(z)\bigg[\bigg(-z+\lambda-\sum_jJ_{ij}J_{ji}\tilde{R}_j(-z)\bigg)\bigg(\sum_{a}K_{ai}^2 \tilde{\hc}_{a}(z)+\sum_{j}J_{ji}^2\tilde{\hc}_{j}(z)\bigg)^{-1}\label{uffa400}\notag\\
&\qquad\quad\bigg(z+\lambda-\sum_jJ_{ij}J_{ji}\tilde{R}_j(z)\bigg)- \bigg(\sum_{j}J_{ij}^2\tilde{C}_j(z)+\Sigma_i\bigg)\bigg]=1\\
&\tilde{\hc}_a(z)\bigg[\Sigma_a+\sum_{j}K_{aj}^2\tilde{C}_j(z)\bigg]=-1\label{uffa5}
\end{align}
\end{subequations}
In \eqref{uffa400} we have substituted the expression for $\tilde{R}_i(z)$ as given by \eqref{uffa2} and used the 
fact that $\tilde{C}_i(z)\bigg(\sum_{j}J_{ij}^2\tilde{C}_j(z)+\Sigma_i\bigg)^{-1}=
\tilde{\hc}_{i}(z)\bigg(\sum_{a}K_{ai}^2 \tilde{\hc}_{a}(z)+\sum_{j}J_{ji}^2 \tilde{\hc}_{j}(z)\bigg)^{-1}$, as can be deduced by comparison
of \eqref{intc1} and \eqref{intb1}.
\section{Quantitative tests}
\label{quant_tests}
In this final section we assess how the extended Plefka method performs
for finite-size systems, by comparing it to the exact solution. For our linear (Gaussian) model we have the advantage of not requiring sampling estimates 
for the  latter, as it can be computed by application of the Kalman filter \cite{kalmanor}.

\subsection{Equal time posterior variance}
\label{quant_tests_postvar}
Specifically, we focus on the computation of posterior variances, since the posterior mean given by the extended Plefka expansion
coincides with the exact result from the Kalman filter (see section \ref{sec:pwoPM}). 
The Kalman filter consists of a forward and backward iteration, starting respectively from 
one of the two time boundaries $t=0$ and $t=T$; away from these, i.e.\ for $0 \ll t \ll T$, the forward-backward recursion converges to
stationary values and this is the regime we concentrate on. The exact expressions for the posterior variance in this stationary regime expressions are well known in filtering theory \cite{vinter}; 
for reference we summarize them here, in the formulation of \cite{bravikalman}.

We first define shorthands for the coupling matrices: let $\bm{K}^{\rm bb}$ be the $N^{\rm b}\times N^{\rm b}$ matrix with entries $-\lambda \delta_{ij}+ J_{ij}$ 
and $\bm{K}^{\rm sb}$ the $N^{\rm s}\times N^{\rm b}$ matrix with entries $K_{ai}$, with a similar definition for $\bm{K}^{\rm bs}$. We denote the \emph{exact} equal time posterior variance 
as $\bm{C}^{*}(0)$; it satisfies a Lyapunov equation \cite{bravikalman} 
\begin{equation}
\label{eq:post}
 \bm{K}^{\rm bb|s}\bm{C}^{*}(0)+ \bm{C}^{*}(0)\bm{K}^{\rm bb|s\,\it{T}}+\sigma_{\rm b}^2=0
\end{equation}
with $\ldots^T$ denoting matrix transpose. $\bm{K}^{\rm bb|s}$ is the bulk-bulk coupling matrix conditioned on the observed 
subnetwork trajectory, as the superscript $\rm bb|s$ indicates. We can think of it as a ``posterior'' drift,
which incorporates the effect of observations via a matrix $\bm{A}$, as follows
\be
\label{eq:postdrift}
\bm{K}^{\rm bb|s}= \bm{K}^{\rm bb} - \sigma_{\rm b}^2\bm{A}
\ee
The matrix $\bm{A}$ itself solves a Riccati equation
\begin{equation}
\label{eq:riccati1}
\sigma_{\rm b}^2\bm{A}^2-\bm{A}\bm{K}^{\rm bb}-\bm{K}^{\rm bb\,\it{T}}\bm{A}=\bm{W}
\end{equation}
and can be shown \cite{bravikalman} to be directly related to the backward propagation in the Kalman filter.
$\bm{W}=\bm{K}^{\rm sb\,\it{T}}\bm{K}^{\rm sb}/\sigma_{\rm s}^2$ in \eqref{eq:riccati1} is known as the \emph{feedback}
matrix and describes how observations affect the posterior statistics. Note that we have chosen scalar noise covariances $\Sigma_i = \sigma_{\rm b}^2$ and $\Sigma_a = \sigma_{\rm s}^2$
to fully focus on the interplay of different types of interactions (hidden-hidden, hidden-to-observed) 
in determining the performance of this inference framework.  The two-time stationary posterior variance can be written as an exponential decay with the posterior drift matrix $\bm{K}^{\rm bb|s}$, i.e.\ 
\begin{equation}
\label{postdecay}
\bm{C}^{*}(t)= e^{\bm{K}^{\rm bb|s}t}\bm{C}^{*}(0)
\end{equation}
for positive time lags $t$. 

Given these expressions, we can now compare the \emph{exact} equal time posterior variance
with the \emph{approximated} one via the extended Plefka expansion.
By construction, this approximation allows one to estimate only \emph{single-site} equal time posterior variances $C_i(0)$; each can be found as the numerical inverse Fourier transform
of $\tilde{C}_i(\omega)$ (corresponding to the solution of \eqref{eqs0} with $z =\text{i}\omega$)
\be
\label{ci0}
C_i(0)= \frac{1}{2\pi}\int_{-\infty}^{+\infty}d \omega \,\tilde{C}_i(\omega)
\ee

\subsection{Relaxation times}
\label{times}
In addition to the posterior variances one can define ``posterior'' relaxation times that capture the timescale over which posterior correlations decay. We define these 
as mean-squared decay times
\be
\label{taup}
\tau_i^2=\frac{\int_{-\infty}^{+\infty} t^2 C_i(t)dt}{2\, \tilde{C}_i(0)}
=-\frac{1}{2\, \tilde{C}_i(0)}\frac{d^2 \tilde{C}_i(\omega)}{d^2\omega}\bigg\lvert_{\omega=0}
\ee
Thus $\tau_i$ can be found immediately from the second derivative of $\tilde{C}_i(\omega)$.
The exact (mean-square) relaxation times, denoted here by $\bm{\tau}^{*}$, can be calculated as 
\be
\label{tauex}
\tau_i^{*\,2}= \frac{\int_{-\infty}^{+\infty} t^2 C_{ii}^{*}(t)dt}{2\,\int_{-\infty}^{+\infty} C_{ii}^{*}(t)dt}=
\frac{\sum_c \lambda_c^{-3} (\bm{r}^c \bm{l}^{c}{}^{T}\bm{C}^{*}(0))_{ii}}{\sum_{c} \lambda_{c}^{-1} (\bm{r}^{c}\bm{l}^{c}{}^{T}\bm{C}^{*}(0))_{ii}}
\ee
where $\bm{C}^{*}(t)$ is expressed as in \eqref{postdecay} and directly integrated. Here we have exploited a decomposition
of the matrix exponential, with $\lambda_c$ denoting the eigenvalues of $-\bm{K}^{\rm bb|s}$ and $\bm{r}^c$ and $\bm{l}^{c}$ respectively 
the corresponding right and left eigenvectors.

\subsection{Gaussian couplings}
\label{coup}
So far no restriction has been placed on the distribution of couplings: the derivation of the method 
does \emph{not per se} need any additional assumptions on the structure of interactions. 

Following a standard choice in the study of mean field models, e.g.\ for spin glasses~\cite{sompolinsky1},
we consider in the following all-to-all, long-ranged and randomly distributed couplings of different symmetry. More precisely, 
we assume as in \cite{bravi} that $\lbrace J_{ij} \rbrace$ is a real matrix belonging 
to the Girko ensemble \cite{girko1}, i.e.\ its elements are independently and randomly distributed Gaussian variables with zero mean and variance 
satisfying
\begin{equation}
\label{jij}
 \langle {J}_{ij} {J}_{ij} \rangle=\frac{j^2}{N^{\rm b}}\qquad \langle {J}_{ji} {J}_{ij} \rangle=\frac{\eta\, j^2}{N^{\rm b}}
\end{equation}
The parameter $\eta\in[-1,1]$ describes the degree to which the matrix $\lbrace J_{ij} \rbrace$ is symmetric. 
Importantly, when $\eta < 1$ we do not have detailed balance so are dealing with \emph{non-equilibrium} dynamics. 
We take $\lbrace K_{ai} \rbrace$ similarly as a random matrix with independent Gaussian entries of zero mean and variance ${j^2}/{N^{\rm b}}$. The amplitude $j$ controls the strength of $\lbrace J_{ij} \rbrace$ (hidden-to-hidden) and $\lbrace K_{ai} \rbrace$ (hidden-to-observed) interactions.

For an infinitely large hidden system and interactions chosen in this way, the extended Plefka expansion, conceived as an effectively 
non-interacting representation of the dynamics, is expected to become exact, as we will discuss in detail elsewhere \cite{plefkaobsmf}. Our aim here, then,
is to assess the deviation of the Plefka-approximated results from the exact ones as a consequence of finite system size and depending on the intensity of couplings. 

\subsection{Results} 
\label{sec:results}
\subsubsection{Variation with $N^{\rm b}$.}
We first explore the dependence of the extended Plefka expansion performance on $N^{\rm b}$, the hidden system size. We show 
scatter plots with the exact single-site (diagonal) posterior variances, $C^{*}_{ii}(0)$, on the horizontal axis and the Plefka-approximated 
ones, $C_{i}(0)$, on the vertical axis. Each random realization of the couplings $J_{ij}$ and $K_{ai}$ gives $N^{\rm b}$ points in such a plot, one per hidden node. We repeat this for $M$ realizations, with $M$ chosen to get $M N^{\rm b}=500$ points per plot. The other 
parameters are set to $j=\sigma_{\rm b}=\sigma_{\rm s}=1$, $\eta=0.5$, $\lambda=2.5$\footnote{We have chosen $\lambda=\lambda_{\text{crit}}+1$, 
where $\lambda_{\text{crit}}$ is the lowest value that guarantees
the stability of the hidden dynamics. When we vary $j$ away from unity as in section \ref{coup}, we maintain $\lambda = \lambda_{\text{crit}}+1$ 
where now $\lambda_{\text{crit}}=j(1+\eta)$.}. 
 
\begin{figure}
\includegraphics[width=0.48\textwidth]{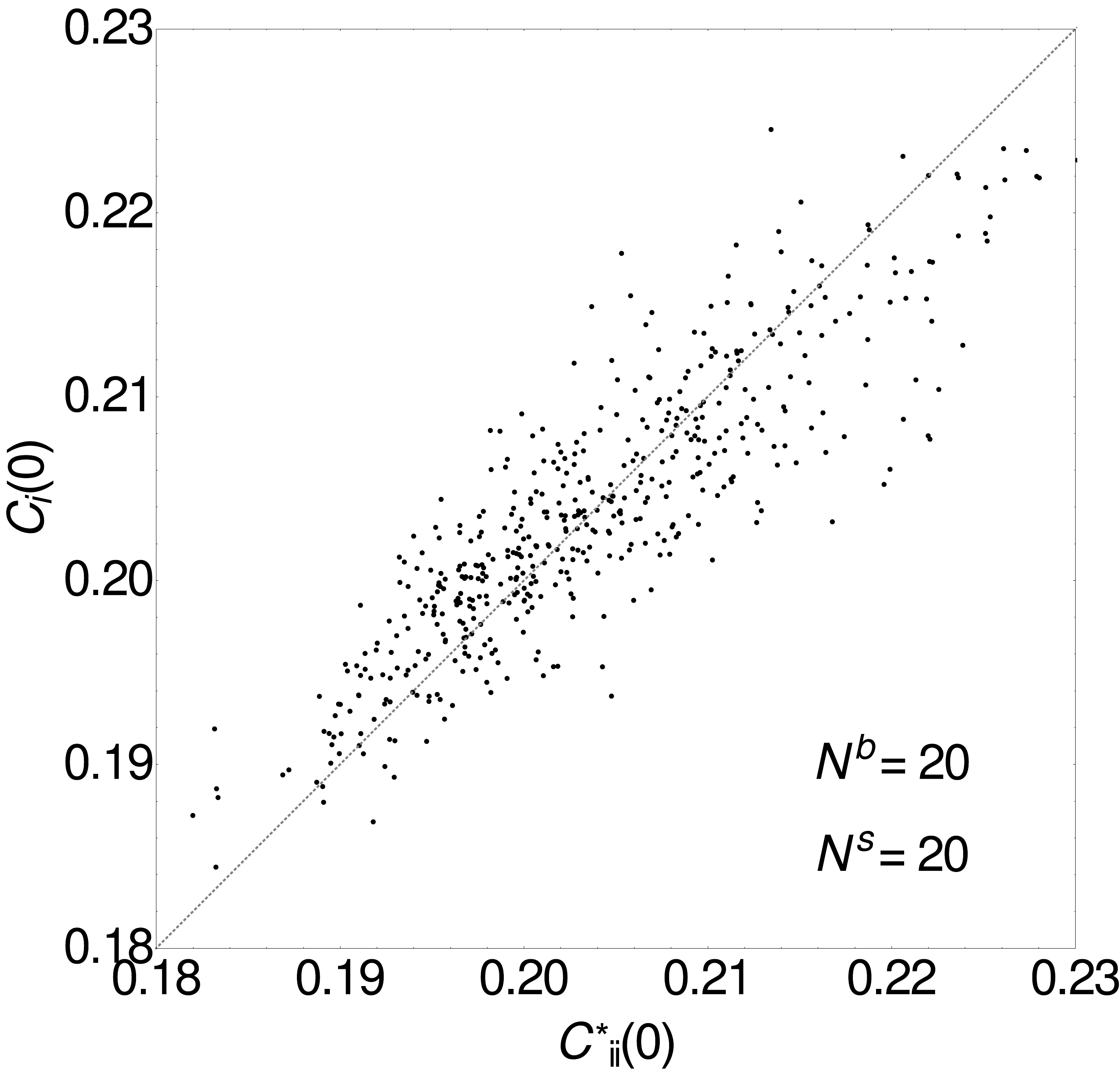}
\includegraphics[width=0.48\textwidth]{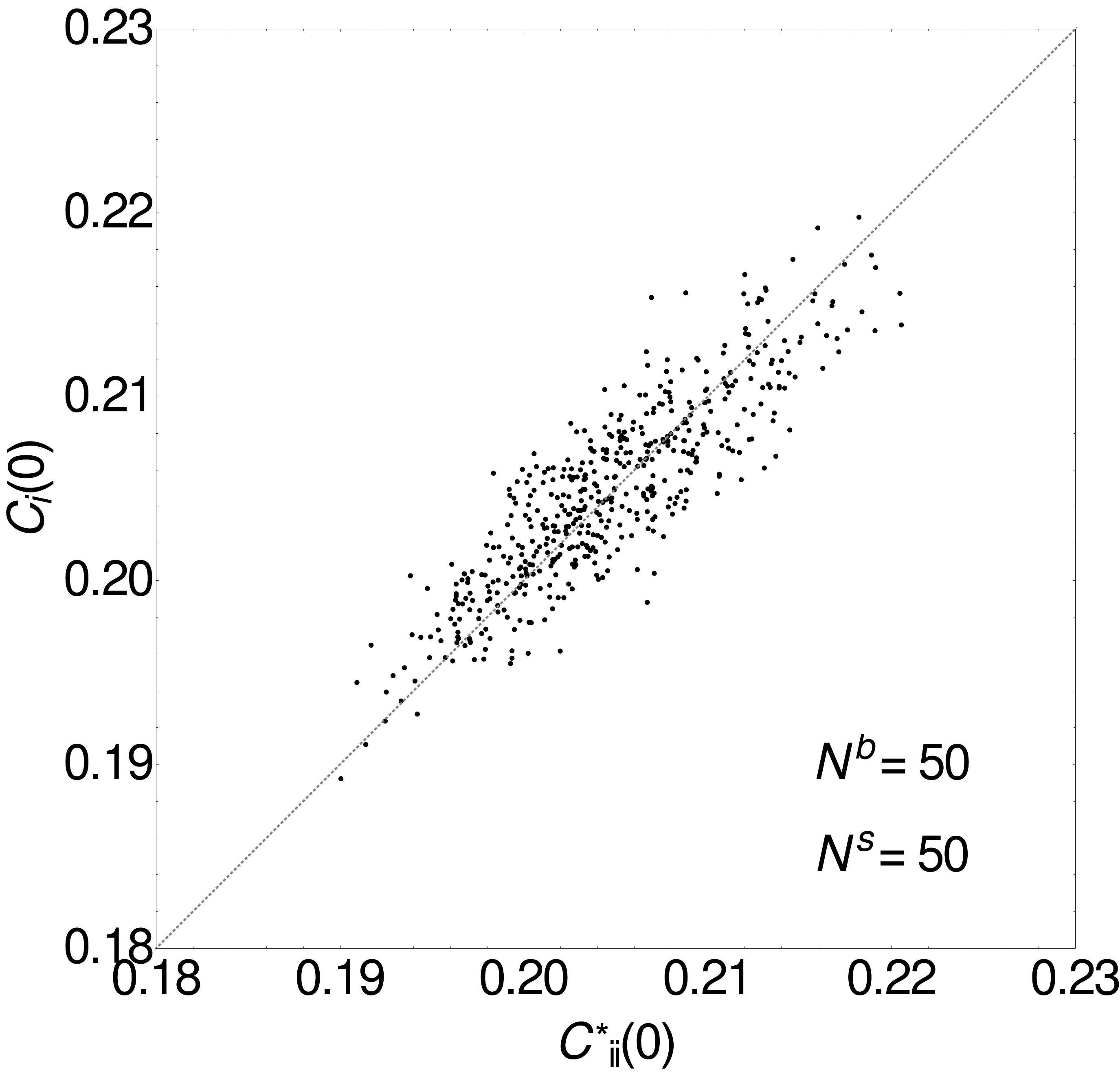}\\
 \caption{Scatter plots of equal time posterior variances, showing Plefka predictions versus exact results,
 at fixed ratio $N^{\rm s}/N^{\rm b}=1$
 for $N^{\rm b}=20$ (left) and for $N^{\rm b}=50$ (right). The diagonal (in light gray) is the reference line of perfect prediction.}
 \label{fig:scatternc}
\end{figure}

Figures \ref{fig:scatternc} and \ref{fig:scatternc0} (left) show data taken
at fixed ratio $N^{\rm s}/N^{\rm b}=1$, with hidden system size $N^{\rm b}$ increasing from 20 to 100. The width of the cloud of points shrinks and all points become more concentrated 
around the center, meaning that the Plefka approximation becomes increasingly close to the exact results for larger hidden systems. 
To quantify this, we use the Mean Square Error ($MSE$) defined as
\be
\label{eq:mse}
MSE=\frac{1}{N^{\rm b}}\sum_{i=1}^{N^{\rm b}}\bigg(C_i(0) - C^{*}_{ii}(0)\bigg)^2
\ee
and plot its average over the $M$ samples as a function of $N^{\rm b}$ in figure \ref{fig:mse} (right): one clearly sees 
a $1/N^{\rm b}$ decay with $N^{\rm b}$. This is expected from the structure of interactions, which are weak and long-ranged (see section \ref{coup}):
for larger $N^{\rm b}$ the dynamical behaviour becomes increasingly mean-field, thus the Plefka-approximated values resemble more closely the exact ones.

The variance in the scatter plots along both axes is a measure of the heterogeneity in the posterior across nodes, taking into account also different interaction realizations. 
We denote these variances $\text{Var}(\bm{C}^{*})$ for the exact (horizontal) axis and
$\text{Var}(\bm{C})$ for the Plefka predictions. These two variances also
decrease with $N^{\rm b}$, as is clear by eye from the trend across figures \ref{fig:scatternc} and \ref{fig:scatternc0} (left). 
Plotting the variances against $N^{\rm b}$, as done in figure \ref{fig:mse} (left), shows that they scale as $1/N^{\rm b}$ as would be expected from the fact 
that the posterior variances become self-averaging in the limit $N^{\rm b}\to \infty$ (see \cite{plefkaobsmf}).

\subsubsection{Variation with $N^{\rm s}$.}

\begin{figure}
\includegraphics[width=0.48\textwidth]{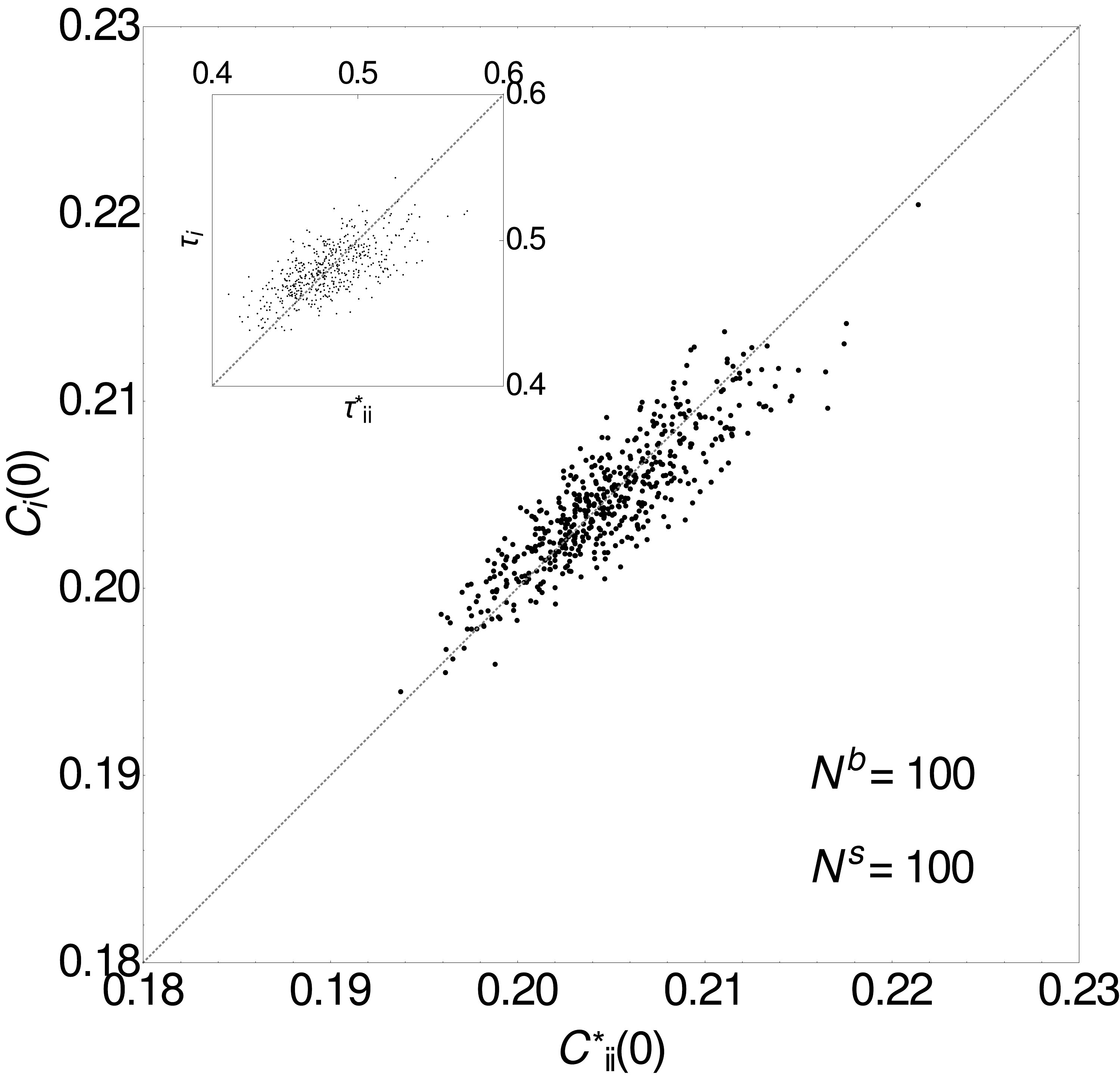}
\includegraphics[width=0.48\textwidth]{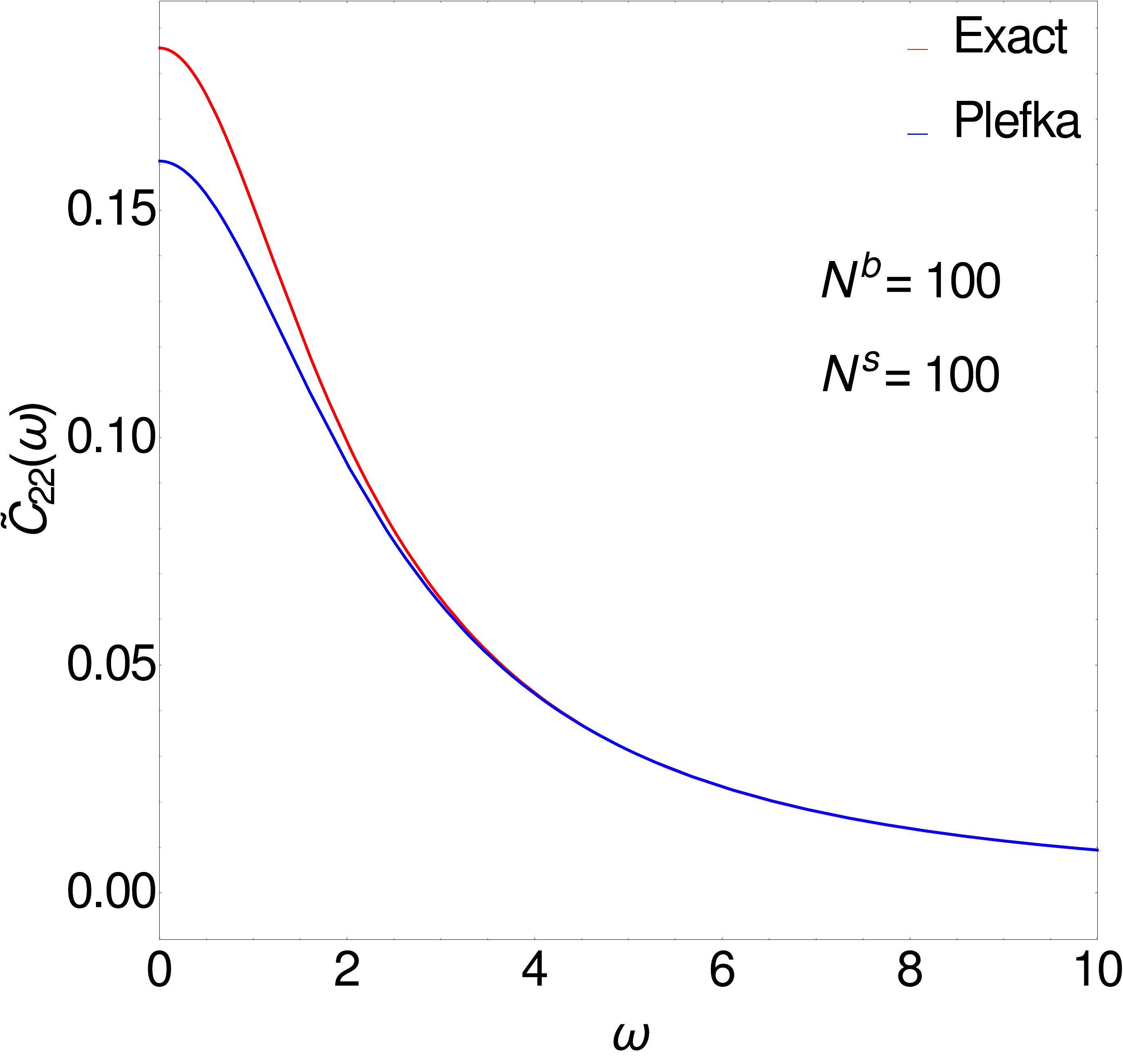}
 \caption{(Left) Scatter plot of equal time posterior variances for $N^{\rm b}=100$, at ratio $N^{\rm s}/N^{\rm b}=1$. 
 The inset shows the scatter plot of posterior relaxation times for the same $N^{\rm b}$ and $N^{\rm s}$.
 (Right) Example of power spectrum for one node ($i=2$) and one sample of $J_{ij}$ and $K_{ai}$ taken from the data on the left.}
 \label{fig:scatternc0}
\end{figure}

\begin{figure}
\includegraphics[width=0.51\textwidth]{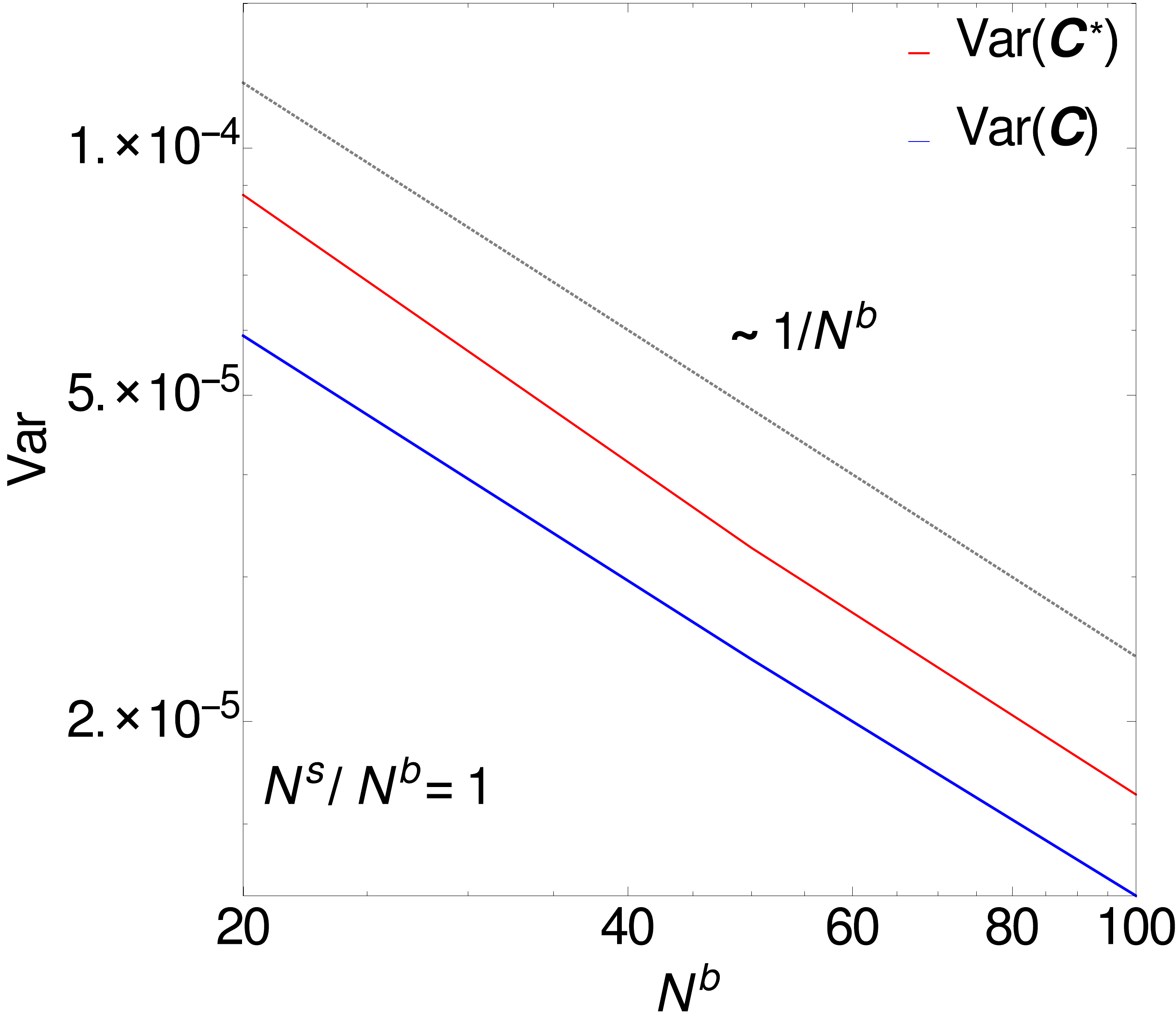}
\includegraphics[width=0.51\textwidth]{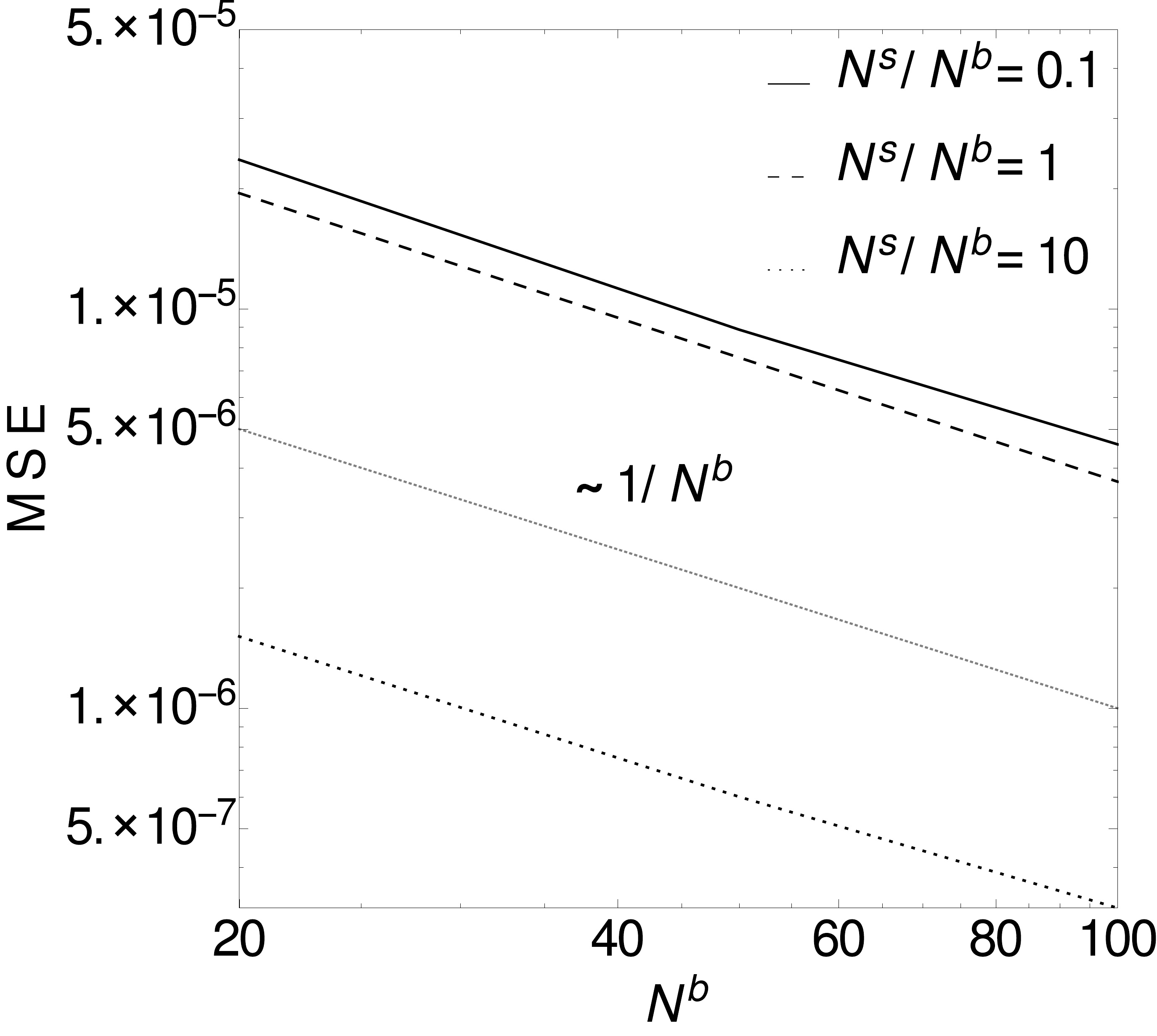}
 \caption{(Left) Variances of the $500$-point samples shown in Figs.\ \ref{fig:scatternc} and \ref{fig:scatternc0} (left) as a function of $N^{\rm b}$, for exact and Plefka predictions.
 (Right)  $MSE$, defined by \eqref{eq:mse}, 
for the data in Figs.\ \ref{fig:scatternc} and \ref{fig:scatternc0} (left), as a function of $N^{\rm b}$ .
 Both plots are on log-log axes; the light gray line shows a $1/N^{\rm b}$ power law as a guide to the eye.}
 \label{fig:mse}
\end{figure}

\begin{figure}
\includegraphics[width=0.48\textwidth]{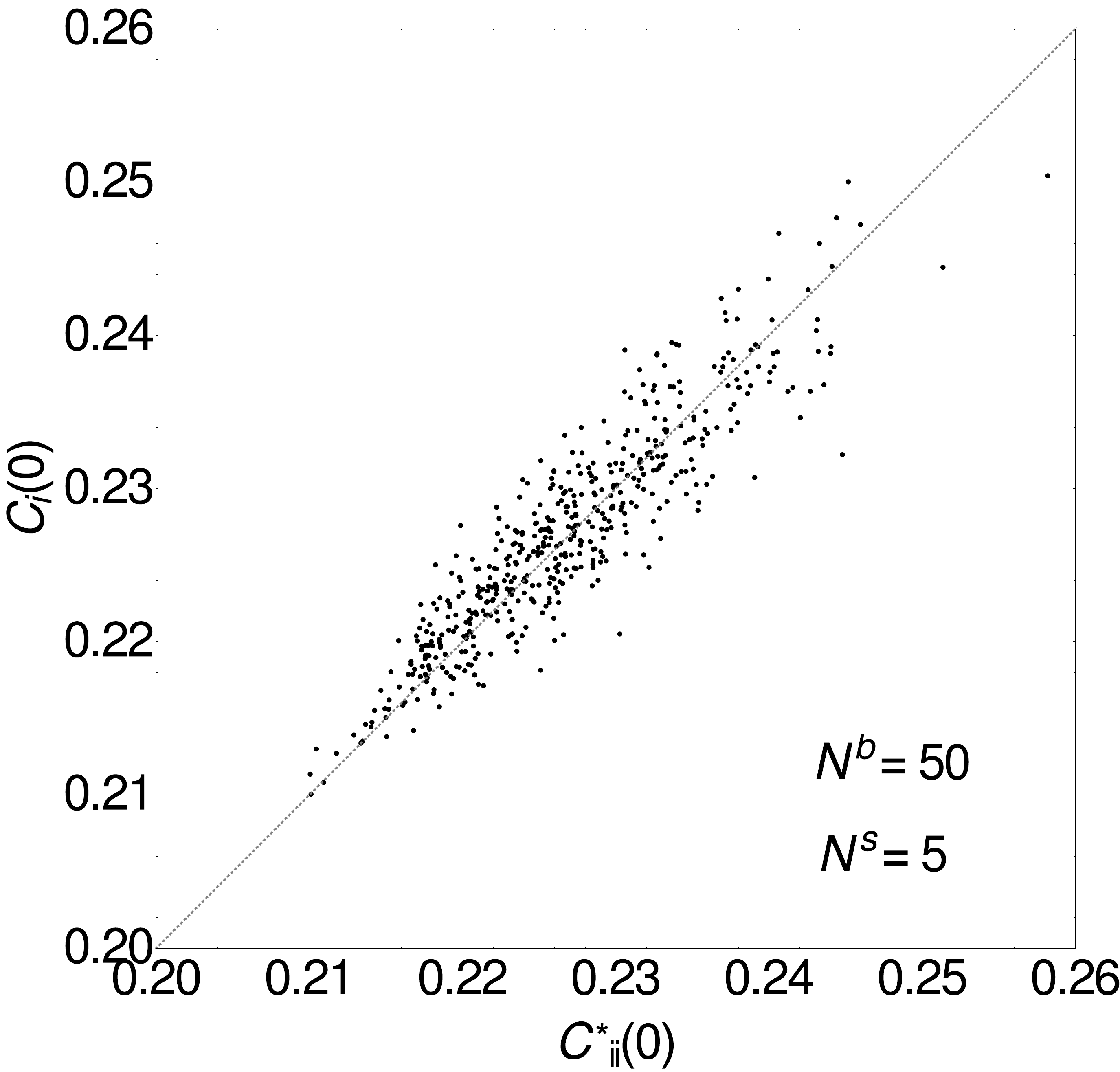}
\includegraphics[width=0.48\textwidth]{corra1n50-eps-converted-to.pdf}\\
\includegraphics[width=0.48\textwidth]{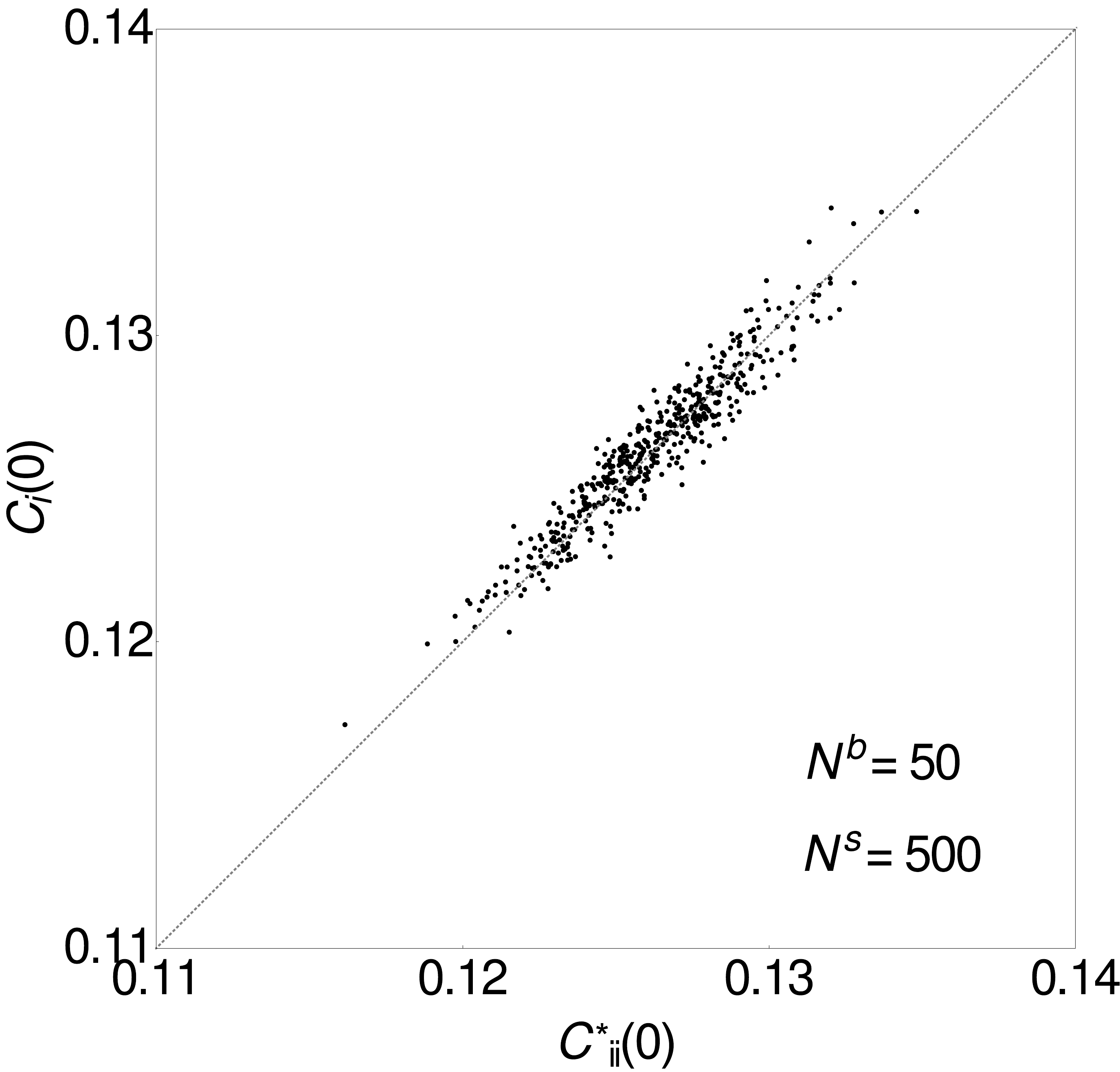}
\includegraphics[width=0.49\textwidth]{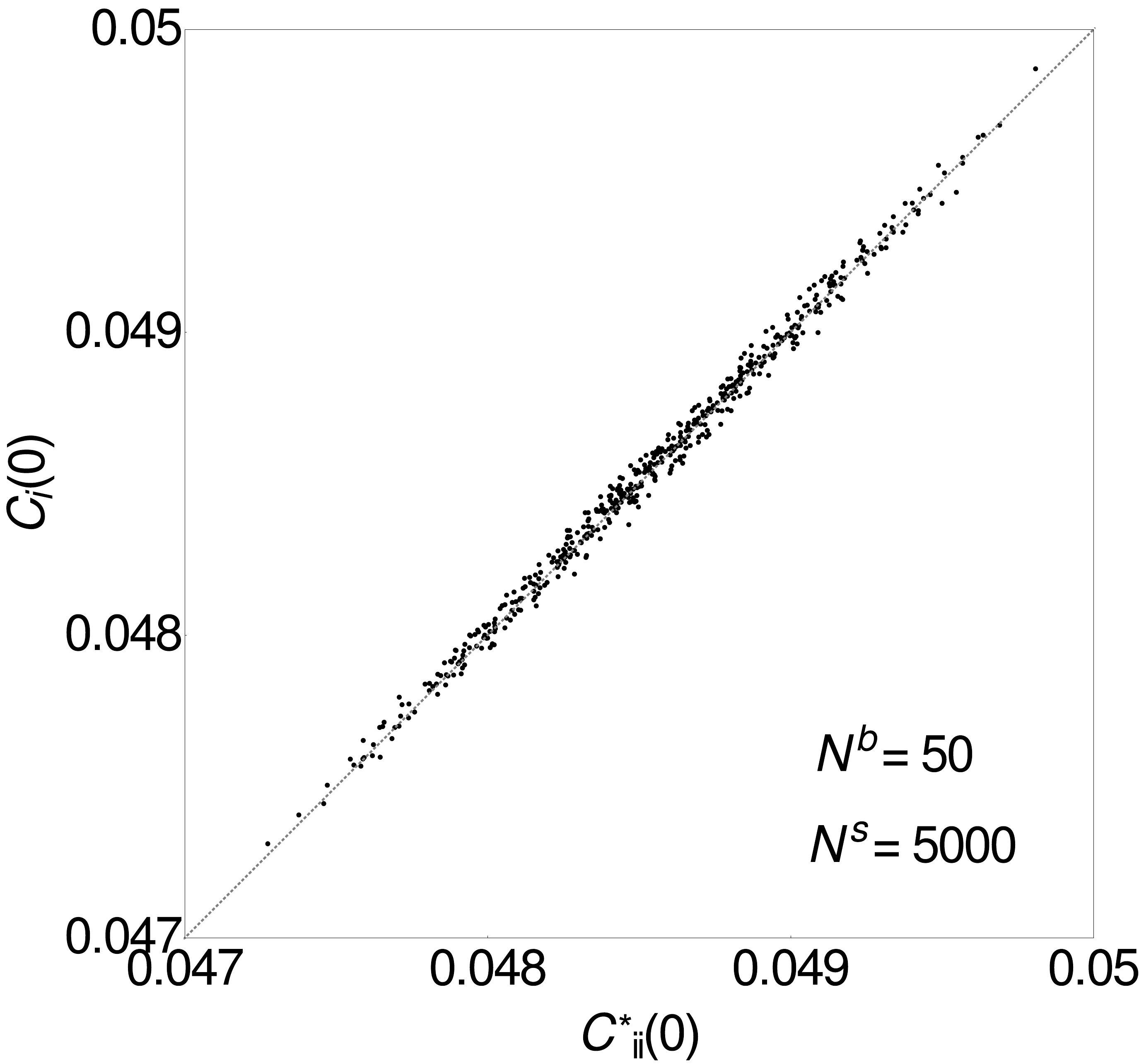}
 \caption{Scatter plots of equal time posterior variances,
 at fixed $N^{\rm b}=50$ and for increasing ratio $N^{\rm s}/N^{\rm b}=0.1$ (left top), $1$ (right top), $10$ (left bottom) and $100$ (right bottom).}
 \label{fig:scatterac}
\end{figure}

Closer inspection of figure \ref{fig:mse} (left) shows
that, at fixed ratio $N^{\rm s}/N^{\rm b}=1$, the variance of the exact predictions is systematically larger than for the Plefka expansion, meaning that the Plefka expansion does 
not completely reproduce the heterogeneity within and across networks. In Figs.\ \ref{fig:scatternc} and \ref{fig:scatternc0} (left), the same effect shows up in the slope of the point cloud, which lies below $1$ (diagonal line). We now investigate how this behaviour changes with $N^{\rm s}$, the number of observed nodes.
In figure \ref{fig:scatterac} we display scatter plots for fixed $N^{\rm b}=50$ and increasing $N^{\rm s}$. The points tend to align along the diagonal and the slope of the cloud approaches 1, implying that the Plefka predictions become increasingly good at capturing the true heterogeneity in the system.
The intuition here is that heterogeneity comes from different ``directions" of the dynamics of the hidden nodes being weakly or strongly constrained by observations. This variability is stronger when $N^{\rm s}$ is small, in particular when $N^{\rm s}<N^{\rm b}$ so that some directions are essentially unconstrained \cite{bravikalman}.
Mathematically, the variability manifests itself as a larger spread of eigenvalues in the posterior covariance matrix
$\bm{C}^{*}(0)$. This can be confirmed by inspection of the structure of the exact equations
in section \ref{quant_tests_postvar} for large $N^{\rm s}$ and $N^{\rm b}$.
For large ratios $N^{\rm s}/N^{\rm b}$, the spectrum of the feedback matrix $\bm{W}$ in \eqref{eq:riccati1}, which follows a \emph{Mar$\breve{c}$enko-Pastur} law \cite{pastur}, becomes a narrow peak
(see \cite{bravikalman} for a more detailed spectral analysis). From equation \eqref{eq:riccati1} this will tend to make $\bm{A}$ closer to diagonal, and hence also the posterior 
drift \eqref{eq:postdrift}. The exact posterior variance thus becomes more similar to the one of a non-interacting dynamics, and is therefore better predicted by a mean field scheme such as the extended Plefka expansion.

Together with the variances of the scatter plots, also the $MSE$ decreases with increasing $N^{\rm s}/N^{\rm b}$, as shown in figure \ref{fig:mse} (right). This 
trend of course is partly a consequence of the reduced variation in the $C_{ii}^{*}(0)$ themselves: to understand how the $MSE$ depends on $N^{\rm s}/N^{\rm b}$ if we factor out this underlying effect
we can normalize it by the variance and define
\be
\label{eq:ec}
\epsilon=\frac{MSE}{\text{Var}(\bm{C}^{*})}
\ee
Figure \ref{fig:jey} (left) shows $\epsilon$ for $N^{\rm b}=50$ and different $N^{\rm s}/N^{\rm b}$: at high $N^{\rm s}$, the deviation between the approximate and exact results is very small 
even when we measure it with respect to the underlying variation $\text{Var}(\bm{C}^{*})$.

A separate trend that can be noted from the scatter plots in figure \ref{fig:scatterac} is the fact that, 
for increasing $N^{\rm s}$, the $C_{ii}^{*}(0)$ (and their Plefka equivalents) become smaller in absolute terms. This is reasonable:
by adding constraints to the estimation of the hidden dynamics, the inference errors diminish so that the inferred means become more accurate 
(see also \cite{plefkaobsmf} and \cite{bravikalman}).

Finally we comment briefly on the estimation of posterior relaxation times as defined in section \ref{times}.
The inset of figure \ref{fig:scatternc0} (left) shows an example of a scatter plot. The tilt of the point cloud away from the diagonal is more pronounced than for the posterior variances, indicating that the Plefka approximation performs somewhat worse in predicting timescales. The reason can be seen by comparing the Plefka power spectra to the exact ones. Figure \ref{fig:scatternc0} (right) shows a typical example: the Plefka approximation works very well in the higher-frequency part of the spectrum. The deviations at lower frequencies make relatively little difference to the area under the spectrum, which gives the (equal-time) posterior variance. The curvature at zero frequency, on the other hand, which is crucial in determining the typical time \eqref{taup}, can be affected rather more strongly.

\subsubsection{Variation with $j$ and $\eta$.}  
\begin{figure}
\includegraphics[width=0.48\textwidth]{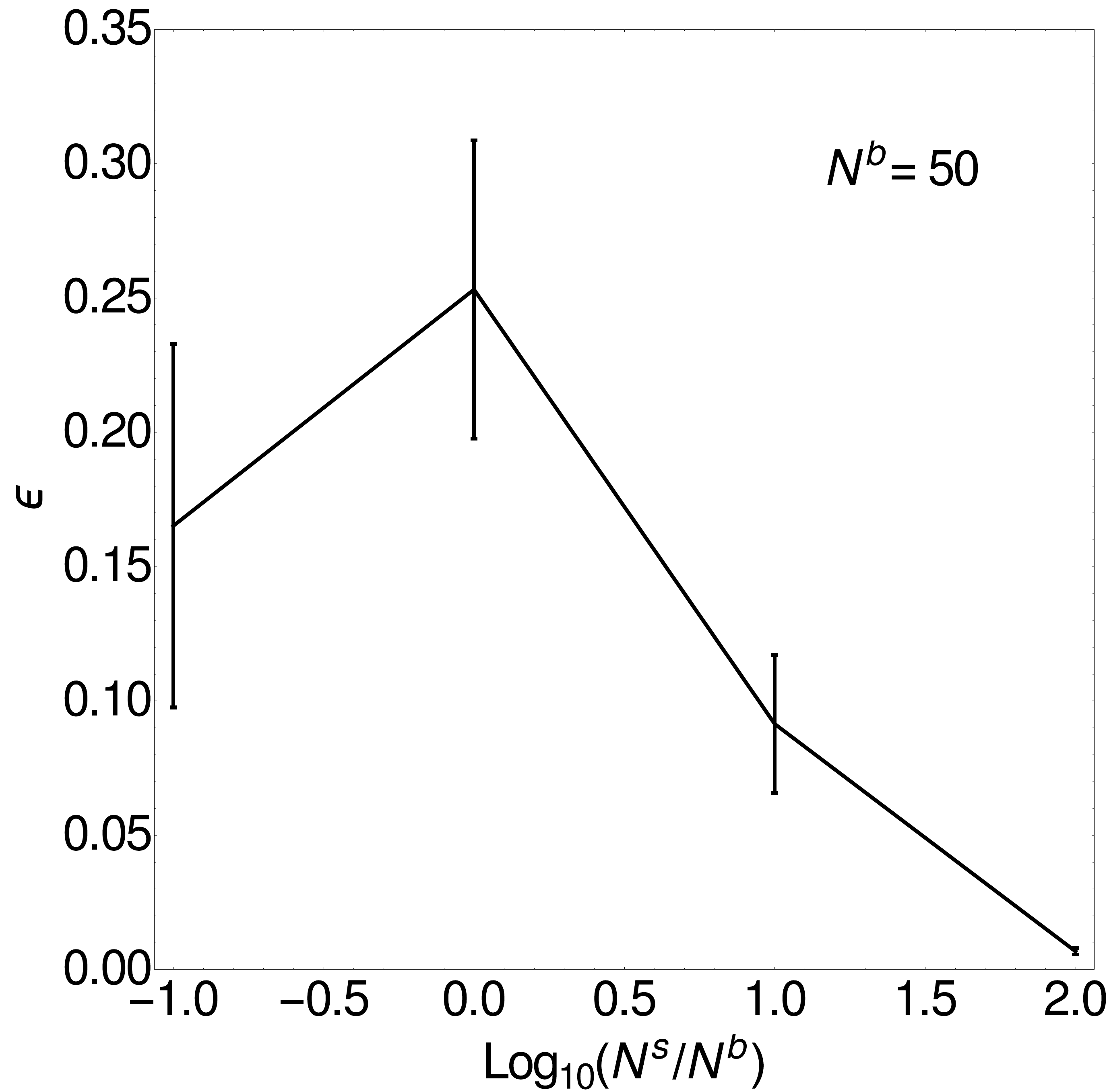}
\includegraphics[width=0.47\textwidth]{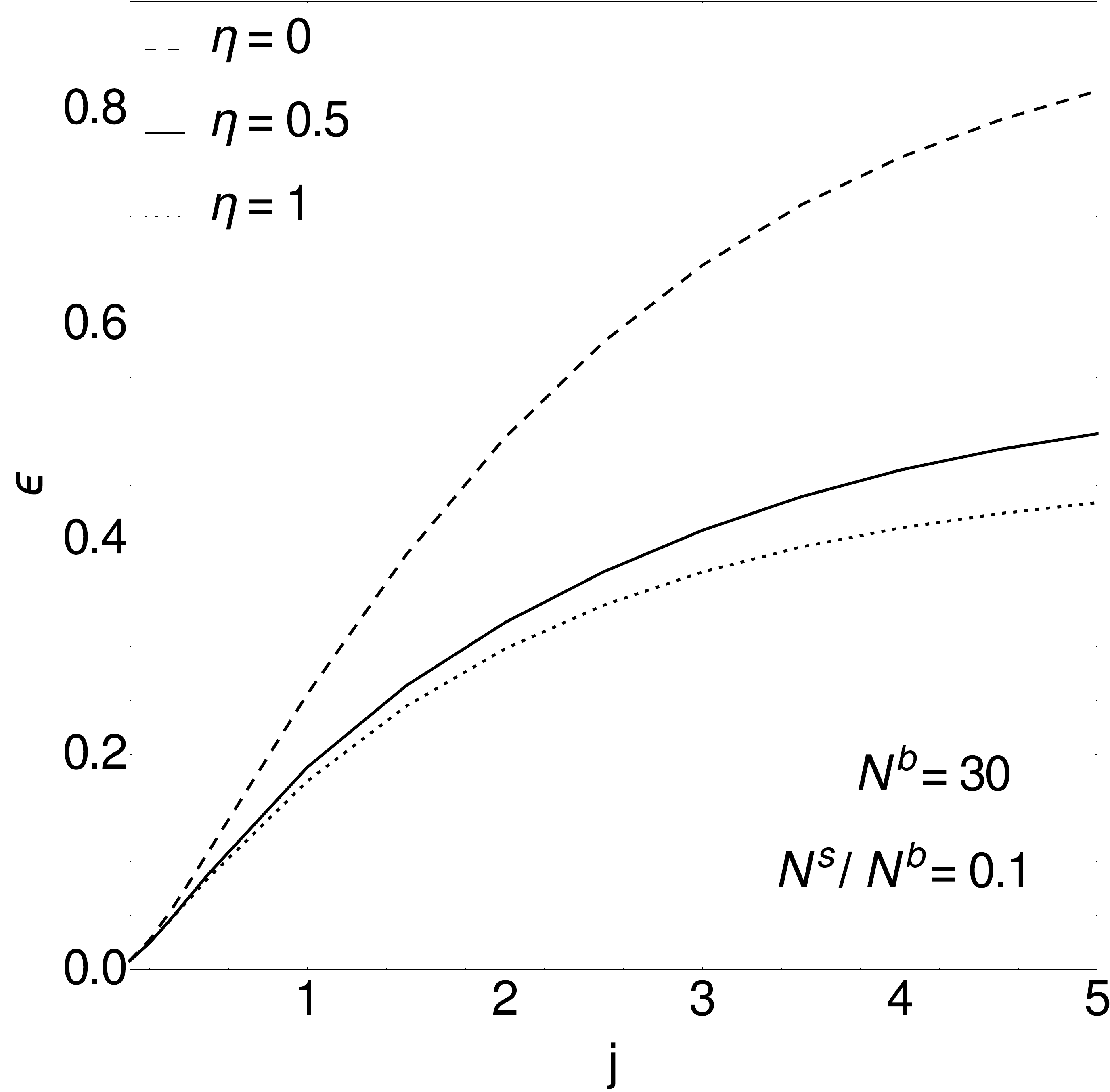}
 \caption{(Left) Variance-normalized error $\epsilon$, as defined by \eqref{eq:ec}, for $N^{\rm b}=50$ and averaged over $M$ coupling realizations, 
as a function of the ratio $N^{\rm s}/N^{\rm b}$. The error bars show the standard deviation across realizations, which diminishes
with increasing $N^{\rm s}$. (Right) $\epsilon$ for $N^{\rm b}=30$ and $N^{\rm s}/N^{\rm b}=0.1$ as a function of $j$, averaged over $M=10$ realizations and for 
different values of the symmetry parameter $\eta$. For every value of $j$ and $\eta$, we set $\lambda=j(1+\eta)+1$, while the noise is chosen as
 $\sigma_{\rm b}=\sigma_{\rm s}=1$.}
 \label{fig:jey}
\end{figure}

The Plefka approach is constructed as a weak coupling expansion, therefore its performance would be expected to deteriorate with higher $j$, leading to
an increasing $MSE$. At fixed $N^{\rm b}$, \eqref{jij} shows that $j$ controls the heterogeneity within the graph, and so the variation across 
 the $C_{ii}^*(0)$ should also increase $j$. To see how these trends combine we plot in figure \ref{fig:jey} (right) the variance-normalized mean square error 
 $\epsilon$ \eqref{eq:ec} against the coupling strength $j$. This increases, showing that the increasing $MSE$ dominates over the increase in spread of the posterior variances.

In figure \ref{fig:jey} (right) we also explore the role of the interaction symmetry parameter $\eta$: one sees that the Plefka predictions get better the closer the system dynamics 
is to equilibrium. This is a non-trivial result, and not simple to explain. The symmetry parameter $\eta$ determines the typical size of 
the ``there and back'' interaction products $J_{ij}J_{ji}$ and these in turn (see \eqref{eqs0}) govern the strength of the response function terms 
in the Plefka treatment. One might therefore speculate that the Plefka does a good job of capturing these response effects, and therefore performs well when they are dominant.

\subsubsection{Computation time.}
\label{sec:cpu}
\begin{figure}
$\qquad\qquad$\includegraphics[width=0.78\textwidth]{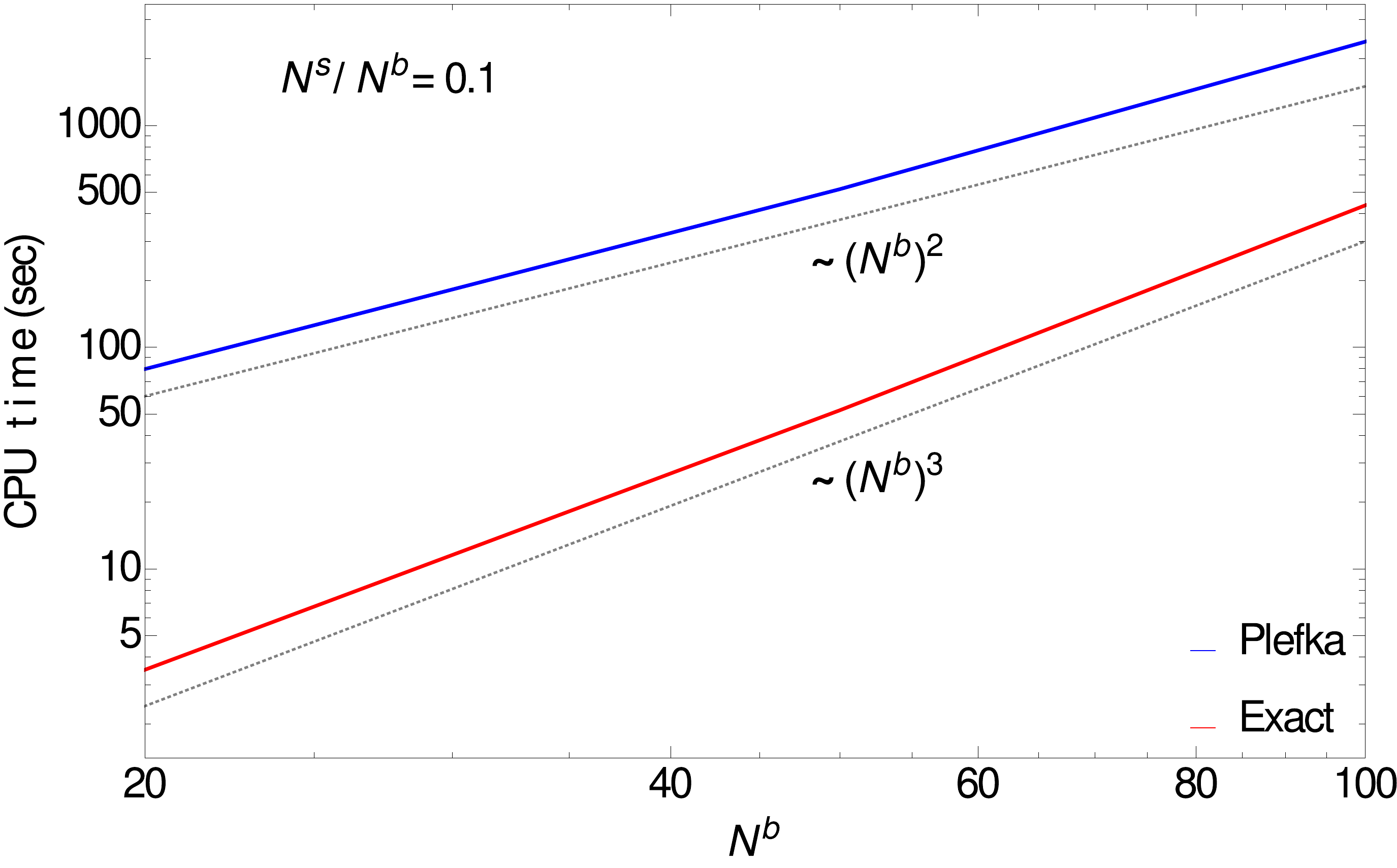}
 \caption{Log-log plot of CPU time (in seconds) for $N^{\rm s}/N^{\rm b}=0.1$ as a function of $N^{\rm b}$ (averaged over the corresponding $M$ realizations). The light gray lines
 show the $\sim (N^{\rm b})^3$ and $\sim (N^{\rm b})^2$ increases for the exact and Plefka computations, respectively.}
 \label{fig:cpu}
\end{figure}

We briefly discuss computational aspects.
In our implementation of the Plefka method, we consider the system \eqref{eqs0} in Fourier space (with $z=\text{i}\omega$). At fixed $\omega$ we solve the coupled equations for the Fourier transforms at each node iteratively, 
setting as stopping criterion $\text{Max}_{i}|\tilde{C}_i(\omega)-\tilde{C}^{'}_i(\omega)|<10^{-8}$, $\tilde{C}^{'}_i$ 
being the value at the previous iteration. Thus the computation time scales as $N_I N_{\Omega} (N^{\rm b})^2$, where $N_I$ is the number of iterations and $N_{\Omega}$ the number of 
frequencies used to sample the power spectra. The $N^{\rm b}$ dependence arises from the 
matrix multiplications required at each iteration, which for the system \eqref{eqs0} gives  $\mathcal{O}((N^{\rm b})^2)$ operations.
The exact computation on the other hand involves a Riccati equation (see \eqref{eq:riccati1}), as well as a matrix diagonalization (see \eqref{tauex}), which both
require $\mathcal{O}((N^{\rm b})^3)$ operations \cite{benner}. For the moderate hidden system sizes analyzed in this paper, the computation (CPU) time remains higher
for the Plefka approach, as is shown in figure \ref{fig:cpu}. But the more benign system size 
scaling -- quadratic rather than cubic -- will make the Plefka method computationally faster above a certain system size threshold; extrapolating from the figure one would expect this crossover to occur already at $N^{\rm b} \approx 1000$.

\section{Conclusion}

In this paper we considered a network of continuous degrees of freedom where some nodes are hidden and the others observed. In this setting
we discussed an application of the Extended Plefka Expansion to the problem of inferring hidden states over time, given an observed time trajectory of the other nodes. We focussed in particular on the case of a linear dynamics with Langevin noise. This choice was motivated by the fact that the posterior statistics, i.e.\ the statistics of hidden trajectories conditioned on 
observations, is Gaussian and thus fully described by its first and second moments. The posterior means give us the optimal hidden state predictions, and the posterior variances the error bars on these, hence also the expected inference error.

The approximation we make in the Extended Plefka Expansion is to consider the second moments of the posterior as local 
in the degrees of freedom, while maintaining them as functions of two times. Because of this, the method yields effective equations of motion for the hidden variables as decoupled from each other and from the observed variables, but with nontrivial temporal self-correlations through memory integrals (response terms) and coloured noises.
In mathematical terms, the Extended Plefka Expansion gives us a system of coupled equations for the posterior physical correlations, responses and auxiliary correlations, the latter having
the role of implementing the constraints arising from the dynamics of the observed nodes. 
Evaluating the posterior covariances at equal times gives the posterior error, hence the Plefka estimate of the inference error.

Having derived the Plefka equations in general form, we then focussed on systems with weak, long range, random interactions to perform 
numerical tests. It is worth stressing that the approach presented here allows one to derive 
results for rather general interaction scenarios. Specifically it holds for any degree of symmetry of the interactions and is thus applicable to non-equilibrium dynamics, 
for which the detailed balance condition is not satisfied. 
Because of the mean field character of the decoupling we assumed in setting up the extended Plefka expansion, 
one would expect the approximation to become exact for couplings with the statistics considered here, in the limit of a large network.
(We verify this explicitly in \cite{plefkaobsmf} and \cite{bravikalman}.) The focus on linear (Gaussian) models is motivated here by 
the possibility of studying the accuracy of our method in comparison to exact results, for finite system sizes and strong 
couplings; these exact results can be obtained using the Kalman filter (and smoother) algorithm
\cite{kalmanor, bishop}. We compared in particular the posterior variances, i.e.\ the diagonal elements of the equal time hidden-to-hidden covariance matrix. Using scatter plots and 
the mean-squared prediction error $MSE$, as an absolute measure or normalized by the spread of posterior variances,
we assessed how the Plefka performance depends on the hidden system size $N^{\rm b}$, the number of observations $N^{\rm s}$
and the coupling strength $j$. As expected for a mean field type approximation applied to fully connected systems with long-ranged Gaussian interactions, 
the agreement improves for large system sizes and for moderate coupling strengths (below $j$ of order unity in our tests). Quantitative tests showed good performance already for modest system sizes ($N^{\rm b}\approx 50,100$), however, which seems promising for applications of the approach to many other possible inference problems.
The number of observed nodes also has a non-trivial effect on the comparison between Plefka results and exact predictions: increasing $N^{\rm s}$ makes the spectrum of posterior variances
more homogeneous and thus more mean field-like, which makes the Plefka approach perform better.
As a conclusion, the Plefka provides an accurate mean field (i.e.\ tailored to single-site quantities)
description for hidden state inference, consistent with findings for an application of the method to the kinetic Ising
model \cite{romanobattistin}. This prediction accuracy 
is obtained by taking into account the whole temporal trajectory via response terms. To capture these one has to work with two-time quantities or, as we did here, in Fourier space. 
This does increase the computational effort. At the same time the Plefka method shows a better scaling of computation time with system size 
$N^{\rm b}$ as discussed in section \ref{sec:cpu}. Beyond system sizes of $N^{\rm b} \approx 1000$ it will therefore be advantageous also in computational terms.
Importantly, the extended Plefka expansion is also generalizable to \emph{arbitrary nonlinearities} (see \cite{bravi} and \ref{appendix:a}), for which the exact posterior 
statistics would generally be computable only by costly direct simulations.

There are many other directions of future and further investigation.
First, in a separate paper \cite{plefkaobsmf} we fully develop the mean field limit expressions for this problem 
to characterize the \emph{average} inference error and its dependence on key system parameters
(such as the number of observed nodes and the interaction strength). The effects of the degree of symmetry $\eta$ prove particularly rich, with interesting 
crossover in the power laws in the vicinity of the equilibrium case $\eta=1$.

In addition, we aim to assess the performance of the Extended Plefka Expansion
in the presence of nonlinearities in the dynamics and on different types of graphs. In \cite{bravi} and here 
we considered fully connected networks described by Gaussian, long-ranged interactions, whose
behaviour is properly mean field. For other network types, e.g.\ Erd\H{o}s-R\'enyi graphs \cite{ER} or more heterogeneous power-law networks, 
it will be interesting to see how accurately the Plefka approach can predict hidden state statistics. One might expect that, as long as typical node degrees are not too small, 
the Gaussian decoupling picture underlying our Plefka approach will remain accurate. Variations in node degree will add heterogeneity in node properties that survives even in 
the large system limit, and it remains to be seen how well the Plefka method captures such local effects.

A number of other novel works could start from the data likelihood \eqref{lik} that the Extended Plefka Expansion predicts: this is an approximate dynamical action for the observed nodes, having integrated out the
hidden node dynamics. As such it describes the observed subnetwork dynamics, and a comparison to other approaches, based 
on path integrals as well \cite{gaussianvar} or on projection methods \cite{katy}, should be revealing.

A further use of the Plefka data likelihood would be to learn unknown hidden-to-hidden or observed-to-hidden couplings, 
which is an important statistical modelling problem for dynamical data. Optimizing the data likelihood with respect to 
the couplings would by definition give a maximum-likelihood estimate for these quantities. The relevant learning rules could 
be developed starting from the Plefka equations for fixed couplings.
It would be interesting then to compare with related studies for non-equilibrium Ising spins \cite{roudi1} and for networks with binary visible units and 
continuous-valued hidden ones \cite{tyrcha}, both also relying on mean field dynamical descriptions.
One could tackle interesting questions such as the accuracy of the inferred couplings and the computational efficiency of the iterative algorithms for implementing the learning rules.

Exploiting further the analogy with dynamical learning in neural networks, one could finally think of several additional studies, such as including noise in the observation 
process \cite{krogh,oppersanguivar,oppersangui1,oppersangui2} or investigating finite-size effects analytically \cite{sollich}.

\section*{Acknowledgement}
This work was supported by the Marie Curie Training Network NETADIS (FP7, grant 290038). The authors acknowledge the stimulating research environment provided by the EPSRC Centre for 
Doctoral Training in Cross-Disciplinary Approaches to Non-Equilibrium Systems (CANES, EP/L015854/1). The authors are indebted to Manfred Opper for helpful suggestions and insightful discussions.

\cleardoublepage
\appendix
\section{Generic drift}
\label{appendix:a}
In this Appendix, we provide the perturbative expressions of the Gibbs free energy $G$ and the fields $\bm{h}^{\cdot}$ ($\cdot=\rm s, \rm b$)
for generic interaction terms $\phi_a(\bm{x}^{\rm b}(t), \bm{x}^{\rm s}(t))$ and $\phi_i(\bm{x}^{\rm b}(t), \bm{x}^{\rm s}(t))$.\\
At the first order, $G^1$ is given by \eqref{eq:g1g}, which in this case can be further developed as
\begin{eqnarray}
G^1=&-&\Delta\sum_{it}\bigg( \text{i} \hat{\mu}_i(t)\langle\phi_i(\bm{x}^{\rm b} (t),\bm{x}^{\rm s}(t))\rangle-
\delta R_i(t,t)\bigg\langle \frac{\partial \phi_i(\bm{x}^{\rm b} (t),\bm{x}^{\rm s}(t))}{\partial x_i(t)}\bigg\rangle\bigg)\notag\\
&-&\Delta\sum_{at}\bigg( \text{i} \hat{\mu}_a(t)\langle\phi_a(\bm{x}^{\rm b} (t),\bm{x}^{\rm s}(t))\rangle\bigg)
\label{eq:G1_full_app}
\end{eqnarray}
To find the corresponding fields we apply \eqref{eq:hobs_generic} for the first order
\be
\bm{h}^{1\cdot}=-\frac{1}{\Delta}\frac{\partial G^{1}}{\partial\bm{m}^{\cdot}} \qquad \cdot=\rm s, \rm b
\label{eq:hobs_1}
\ee
which gives explicitly
\begin{subequations}
\begin{align}
\begin{split}
\psi_i^1(t)=&\sum_{j} \bigg(\text{i}\hat{\mu}_{j}(t)
\frac{\partial \langle\phi_j(\bm{x}^{\rm b}(t), \bm{x}^{\rm s}(t))\rangle}{\partial \mu_i(t)}
-\delta R_j(t,t) \frac{\partial}{\partial \mu_i(t)}\bigg\langle \frac{\partial \phi_j(\bm{x}^{\rm b}(t), \bm{x}^{\rm s}(t))}{\partial x_j(t)}\bigg\rangle\bigg)\\
&\quad -\text{i}\hat{\mu}_i(t)\bigg\langle \frac{\partial \phi_i(\bm{x}^{\rm b}(t), \bm{x}^{\rm s}(t))}{\partial x_i(t)}\bigg\rangle
+\sum_a \text{i}\hat{\mu}_a(t)\frac{\partial \langle \phi_a(\bm{x}^{\rm b}(t), \bm{x}^{\rm s}(t))\rangle}{\partial \mu_i(t)}
\end{split}\\
l_i^1(t)=&\quad\mu_i(t)\bigg\langle\frac{\partial \phi_i(\bm{x}^{\rm b}(t), \bm{x}^{\rm s}(t))}{\partial x_i(t)}\bigg\rangle
-\langle\phi_i(\bm{x}^{\rm b}(t), \bm{x}^{\rm s}(t))\rangle\\
l_a^1(t)=&-\langle\phi_a(\bm{x}^{\rm b}(t), \bm{x}^{\rm s}(t))\rangle\\
\hat{R}_i^1(t,t')=&{}-\frac{1}{\Delta}\bigg\langle\frac{\partial \phi_i(\bm{x}^{\rm b}(t), \bm{x}^{\rm s}(t))}{\partial x_i(t)} \bigg\rangle\delta_{tt'}\\
\begin{split}
\hat{C}_i^1(t,t')=
&\quad\frac{1}{\Delta}\sum_{j}\bigg(\text{i}\hat{\mu}_j(t)\frac{\partial\langle \phi_j(\bm{x}^{\rm b} (t),\bm{x}^{\rm s}(t))\rangle}{\partial C_i(t,t)}-
\delta R_j(t,t)\frac{\partial}{\partial C_i(t,t)}\bigg\langle \frac{\partial \phi_j(\bm{x}^{\rm b} (t),\bm{x}^{\rm s}(t))}{\partial x_j(t)}\bigg\rangle\bigg)\delta_{tt'}\\
&+\frac{1}{\Delta}\sum_{a}\bigg(\text{i} \hat{\mu}_a(t)\frac{\partial \langle\phi_a(\bm{x}^{\rm b} (t),\bm{x}^{\rm s}(t))\rangle}{\partial C_i(t,t)}\bigg)\delta_{tt'}
\end{split}\\
\hat{B}_i^1(t,t')=&\hat{B}_a^1(t,t')=0
\end{align}
\label{eq:fieldsnl}
\end{subequations}
For the case of linear dynamics, where $\phi_i(\bm{x}^{\rm b} (t), \bm{x}^{\rm s}(t))=\sum_j J_{ij} x_j(t)+\sum_a K_{ia} x_a(t)$ and 
$\phi_a(\bm{x}^{\rm b}(t) , \bm{x}^{\rm s}(t))=\sum_b J_{ab} x_b(t)+\sum_j K_{aj} x_j(t)$  with 
no self-interactions ($J_{aa}=J_{ii}=0$), one clearly recovers \eqref{eq:fieldsl}.

For the second order, we assume that
\be
\frac{\partial \phi_i(\bm{x}^{\rm b}, \bm{x}^{\rm s})}{\partial x_i}=0
\ee
as this restriction substantially reduces the number of terms in the final result  
(the derivation can nevertheless be carried out for fully general drifts similarly to what has be done in the case without observations,
see \cite{bravi}, Appendix B, and \cite{thesis}). Intuitively, this assumption means that $x_i$ interacts with itself only via the linear term $-\lambda x_i$.
The first order fields \eqref{eq:fieldsnl} can be simplified accordingly
and used in the definition of $d\Xi_{\alpha}/d\alpha$ \eqref{eq:der} to get 
\begin{eqnarray}
\label{eq:dXi_alpha_TAP_d}
\fl\delta\bigg(\frac{d \Xi_{\alpha}}{d \alpha}\bigg)=&&\frac{d \Xi_{\alpha}}{d \alpha} -\bigg\langle \frac{d \Xi_{\alpha}}{d \alpha} \bigg\rangle_0 
=\\
\fl&&-\Delta\sum_{it} \bigg[ \text{i} \delta \hat{x}_i(t)\delta\phi_i(\bm{x}^{\rm b}(t), \bm{x}^{\rm s}(t))
+\text{i} \hat{\mu}_i(t) f_i(\bm{x}^{\rm b}(t), \bm{x}^{\rm s}(t),\bm{\mu}^{\rm b}(t),\bm{C}^{\rm b}(t,t))\bigg]\notag\\
\fl&&-\Delta\sum_{at}
\bigg[ \text{i} \delta \hat{x}_a(t)\delta\phi_a(\bm{x}^{\rm b}(t), \bm{x}^{\rm s}(t))+
\text{i} \hat{\mu}_a(t) f_a(\bm{x}^{\rm b}(t), \bm{x}^{\rm s}(t),\bm{\mu}^{\rm b}(t),\bm{C}^{\rm b}(t,t))\bigg]\notag
\end{eqnarray}
where we have defined
\begin{subequations}
\begin{align}
\begin{split}
&f_i(\bm{x}^{\rm b}(t), \bm{x}^{\rm s}(t),\bm{\mu}^{\rm b}(t),\bm{C}^{\rm b}(t,t))=\\
&\sum_{j}\bigg(\delta \phi_i(\bm{x}^{\rm b}(t), \bm{x}^{\rm s}(t))-\frac{\partial \langle \phi_i(\bm{x}^{\rm b}(t), \bm{x}^{\rm s}(t))\rangle}{\partial \mu_j(t)}\delta x_j(t) - 
\frac{\partial \langle \phi_i(\bm{x}^{\rm b}(t), \bm{x}^{\rm s}(t))\rangle}{\partial C_j(t,t)}\delta (x_j(t)x_j(t))\bigg)
\end{split}\\
\begin{split}
&f_a(\bm{x}^{\rm b}(t), \bm{x}^{\rm s}(t),\bm{\mu}^{\rm b}(t),\bm{C}^{\rm b}(t,t))=\\
&\sum_{j}\bigg(\delta \phi_a(\bm{x}^{\rm b}(t), \bm{x}^{\rm s}(t))-
\frac{\partial \langle \phi_a(\bm{x}^{\rm b}(t), \bm{x}^{\rm s}(t))\rangle}{\partial \mu_j(t)}\delta x_j(t) -
\frac{\partial \langle \phi_a(\bm{x}^{\rm b}(t), \bm{x}^{\rm s}(t))\rangle}{\partial C_j(t,t)}\delta (x_j(t)x_j(t))\bigg)                       
\end{split}
\end{align}
\end{subequations}
and we have denoted by $\bm{\mu}^{\rm b}(t)$ the set $\lbrace \mu_i(t) \rbrace$ and similarly 
$\bm{C}^{\rm b}(t,t) = \lbrace C_i(t,t) \rbrace$. Clearly for the linear dynamics considered in the main text 
$f_i(\bm{x}^{\rm b}, \bm{x}^{\rm s},\bm{\mu}^{\rm b},\bm{C}^{\rm b})=f_a(\bm{x}^{\rm b}, \bm{x}^{\rm s},\bm{\mu}^{\rm b},\bm{C}^{\rm b})\equiv 0$.

As stated in \eqref{eq:g2}, $G^2$ can be found as $\left \langle \left(\delta\left(d \Xi_{\ap}/d \ap\right)\right)^2 \right \rangle_{0}$;
once this is worked out, the fields are given by \eqref{eq:hobs_generic} taken at the second order
\be
\bm{h}^{2\cdot}=-\frac{1}{\Delta^2}\frac{\partial G^{2}}{\partial\bm{m}^{\cdot}} \qquad \cdot=\rm s, \rm b
\label{eq:hobs_2}
\ee
To simplify the algebra, we will assume from now on that the drifts $\phi_i$ (and $\phi_a$) are additive
combinations of functions of the other hidden variables $x_j$, i.e.\ 
of the form $\phi_i(\bm{x}^{\rm b}, \bm{x}^{\rm s}) = \sum_{j\neq i} g _{ij} (x_j,\bm{x}^{\rm s})$ 
(and $\phi_a(\bm{x}^{\rm b}, \bm{x}^{\rm s}) = \sum_{j} g _{aj} (x_j,\bm{x}^{\rm s})$)
in such a way that the cross terms between $\phi$ and $f$ in $G^2$ vanish.
In addition, we exploit that the averages are taken at $\alpha=0$, where there are no correlations between different sites $i$ and $j$. These averages 
can then be further simplified using Wick's theorem as the statistics are Gaussian. The resulting expressions are as follows
if we abbreviate $f_i(\bm{x}^{\rm b}(t), \bm{x}^{\rm s}(t),\bm{\mu}^{\rm b}(t),\bm{C}^{\rm b}(t,t))$ as $f_i(t)$, and 
$\phi_{i}(\bm{x}^{\rm b}(t), \bm{x}^{\rm s}(t))$ as $\phi_i(t)$ (similarly for $f_a(t)$ and $\phi_a(t)$)
\begin{subequations}
\begin{align}
\begin{split}
\psi_i^2(t)= &\quad2\Delta \bigg( \sum_{at'}\bigg\langle \frac{\partial \phi_{a}(t')}{\partial x_i(t')}\, 
\frac{\partial \phi_{a}(t)}{\partial x_i(t)}  \bigg\rangle\delta \hc_a(t,t') \mu_i(t') + \sum_{jt'} \bigg\langle \frac{\partial \phi_{j}(t')}{\partial x_i(t')}\, 
\frac{\partial \phi_{j}(t)}{\partial x_i(t)}  \bigg\rangle \delta \hc_j(t,t') \mu_i(t')\bigg) \\
&-2\Delta\sum_{jt'}\bigg\langle \frac{\partial \phi_{i}(t')}{\partial x_j(t')}\, 
\frac{\partial \phi_{j}(t)}{\partial x_i(t)}  \bigg\rangle
\delta R_j(t',t) \text{i}\hat{\mu}_i(t')\\
&-\Delta \frac{\partial}{\partial\mu_i(t)} \bigg\langle \bigg(\sum_{jt'}\text{i}\hat{\mu}_j(t')f_j(t')+
\sum_{at'}\text{i}\hat{\mu}_a(t')f_a(t')\bigg)^2\bigg\rangle
\end{split}\\
\begin{split}
l_i^2(t)=&-2\Delta\sum_{jt'}\text{i}\hat{\mu}_i(t') \bigg\langle \frac{\partial \phi_{i}(t)}{\partial x_j(t)}\, 
\frac{\partial \phi_{i}(t')}{\partial x_j(t')}\bigg\rangle\delta C_j(t, t')\\
&+2\Delta\sum_{jt'}\bigg\langle \frac{\partial \phi_{i}(t')}{\partial x_j(t')}\, 
\frac{\partial \phi_{j}(t)}{\partial x_i(t)}  \bigg\rangle\mu_i(t')\delta R_j(t',t)\\
&+2 \Delta  \bigg\langle \bigg(\sum_{jt'}\text{i}\hat{\mu}_j(t')f_j(t')+
\sum_{at'}\text{i}\hat{\mu}_a(t')f_a(t')\bigg)f_i(t) \bigg\rangle
\end{split}\\
\begin{split}
l_a^2(t)=&-2\Delta\sum_{at'}\text{i}\hat{\mu}_a(t')\bigg\langle \frac{\partial \phi_{a}(t)}{\partial x_j(t)}\, 
\frac{\partial \phi_{a}(t')}{\partial x_j(t')}  \bigg\rangle\delta C_j(t, t') \\
&+2\Delta\bigg\langle \bigg(\sum_{jt'}\text{i}\hat{\mu}_j(t')f_j(t')+\sum_{bt'}\text{i}\hat{\mu}_b(t')f_b(t')\bigg)f_a(t) \bigg\rangle 
\end{split}\\
\hat{R}_i^2(t,t')=& -2 \sum_{j}\bigg\langle \frac{\partial \phi_{i}(t')}{\partial x_j(t')}\, 
\frac{\partial \phi_{j}(t)}{\partial x_i(t)}  \bigg\rangle\delta R_j(t',t)\\
\begin{split}
\hat{C}_i^2(t,t')=& -\sum_{a}\bigg\langle \frac{\partial \phi_{a}(t)}{\partial x_i(t)}\, 
\frac{\partial \phi_{a}(t')}{\partial x_i(t')}\bigg\rangle \delta \hc_a(t,t') -\sum_{j}\bigg\langle \frac{\partial \phi_{j}(t)}
{\partial x_i(t)}\,\frac{\partial \phi_{j}(t')}{\partial x_i(t')}\bigg\rangle \delta \hc_j(t,t')\\
&-\frac{\partial}{\partial C_i(t)} \bigg\langle \bigg(\sum_{jt'}\text{i}\hat{\mu}_j(t')f_j(t')+
\sum_{at'}\text{i}\hat{\mu}_a(t')f_a(t')\bigg)^2\bigg\rangle
\end{split}\\
\hat{B}_i^2(t,t')=& -\sum_{j}\bigg\langle \frac{\partial \phi_{i}(t)}{\partial x_j(t)}\, 
\frac{\partial \phi_{i}(t')}{\partial x_j(t')}\bigg\rangle \delta C_j(t,t')\\
\hat{B}_a^2(t,t')=& -\sum_{j}\bigg\langle \frac{\partial \phi_{a}(t)}{\partial x_j(t)}\, 
\frac{\partial \phi_{a}(t')}{\partial x_j(t')}  \bigg\rangle \delta C_j(t,t')
\end{align}
\label{eq:fieldsnl2}
\end{subequations}
Finally, one can write the dynamics of the conditional means, 
i.e.\ \eqref{meandyn}, \eqref{meandynhat}, \eqref{meandynhata}, and that of the variances, i.e.\ \eqref{integeq}, in terms of the effective fields
$\bm{h}^{\rm eff\, \cdot}$ ($\cdot=\rm s, \rm b$) defined by \eqref{eq:zerofields}, where one should apply expressions
\eqref{eq:fieldsnl} and \eqref{eq:fieldsnl2} and take the $\Delta \to 0$ limit. Again these effective fields simplify to the 
relevant expressions \eqref{eq:eff2m} in the case of linear interactions.
\cleardoublepage
\section*{References}
\bibliography{./Phdbib}
\end{document}